\pdfoutput=1
\documentclass[12pt,a4paper]{article}

\usepackage{ifthen} 
\newboolean{pdflatex}
\setboolean{pdflatex}{true} 

\newboolean{articletitles}
\setboolean{articletitles}{true} 

\newboolean{uprightparticles}
\setboolean{uprightparticles}{false} 

\def\paperauthors{LHCb collaboration} 
\def\paperasciititle{Measurement of the CKM angle \gamma in $B^{\pm} \to D K^*(892)^{\pm}$ decays} 
\def\papertitle{Measurement of the CKM angle $\gamma$ in $B^{\pm} \to D K^*(892)^{\pm}$ decays} 
\def\paperkeywords{{High Energy Physics}, {LHCb}} 
\def\papercopyright{\the\year\ CERN for the benefit of the LHCb collaboration} 
\def\paperlicence{CC BY 4.0 licence}
\def\paperlicenceurl{https://creativecommons.org/licenses/by/4.0/}

\usepackage{xspace} 
\usepackage{upgreek}

\usepackage[top=1in, bottom=1.25in, left=1in, right=1in]{geometry}

\usepackage{microtype}
\usepackage{lineno}  
\usepackage{xspace} 
\usepackage{caption} 

\usepackage{graphicx}  
\usepackage{color}
\usepackage{colortbl}
\graphicspath{{./figs/}} 

\usepackage{amsmath} 
\usepackage{amssymb}
\usepackage{amsfonts}
\usepackage{upgreek} 

\newcommand*\patchAmsMathEnvironmentForLineno[1]{%
\expandafter\let\csname old#1\expandafter\endcsname\csname #1\endcsname
\expandafter\let\csname oldend#1\expandafter\endcsname\csname
end#1\endcsname
 \renewenvironment{#1}%
   {\linenomath\csname old#1\endcsname}%
   {\csname oldend#1\endcsname\endlinenomath}%
}
\newcommand*\patchBothAmsMathEnvironmentsForLineno[1]{%
  \patchAmsMathEnvironmentForLineno{#1}%
  \patchAmsMathEnvironmentForLineno{#1*}%
}
\AtBeginDocument{%
\patchBothAmsMathEnvironmentsForLineno{equation}%
\patchBothAmsMathEnvironmentsForLineno{align}%
\patchBothAmsMathEnvironmentsForLineno{flalign}%
\patchBothAmsMathEnvironmentsForLineno{alignat}%
\patchBothAmsMathEnvironmentsForLineno{gather}%
\patchBothAmsMathEnvironmentsForLineno{multline}%
\patchBothAmsMathEnvironmentsForLineno{eqnarray}%
}

\usepackage{hyperxmp}

\usepackage[pdftex,
            pdfauthor={\paperauthors},
            pdftitle={\paperasciititle},
            pdfkeywords={\paperkeywords},
            pdfcopyright={Copyright (C) \papercopyright},
            pdflicenseurl={\paperlicenceurl}]{hyperref}

\usepackage[colorinlistoftodos,textsize=scriptsize]{todonotes}

\usepackage[bottom,flushmargin,hang,multiple]{footmisc}

\usepackage[all]{hypcap} 

\usepackage{xspace} 
\usepackage{upgreek}

\def\lhcb   {\mbox{LHCb}\xspace}

\def\besiii {\mbox{BESIII}\xspace}
\def\cleo   {\mbox{CLEO}\xspace}

\def\MagUp {\mbox{\em Mag\kern -0.05em Up}\xspace}

\ifthenelse{\boolean{uprightparticles}}%
{
 
 \def\Pgamma      {\ensuremath{\upgamma}\xspace}

 \def\Ppi         {\ensuremath{\uppi}\xspace}                 
                  
 \def\Prho        {\ensuremath{\uprho}\xspace}

 \def\PDelta      {\ensuremath{\Delta}\xspace}                 
 \def\PXi         {\ensuremath{\Xi}\xspace}                 
 \def\PLambda     {\ensuremath{\Lambda}\xspace}                 
 \def\PSigma      {\ensuremath{\Sigma}\xspace}                 
 \def\POmega      {\ensuremath{\Omega}\xspace}                 
 \def\PUpsilon    {\ensuremath{\Upsilon}\xspace}
 \let\oldPi\Pi
 \def\PPi         {\ensuremath{\oldPi}\xspace}

 \def\PB      {\ensuremath{\mathrm{B}}\xspace}                 
                  
 \def\PD      {\ensuremath{\mathrm{D}}\xspace}

 \def\PK      {\ensuremath{\mathrm{K}}\xspace}

 \def\Pb      {\ensuremath{\mathrm{b}}\xspace}                 
 \def\Pc      {\ensuremath{\mathrm{c}}\xspace}

 \def\Ph      {\ensuremath{\mathrm{h}}\xspace}                 
 \def\Pi      {\ensuremath{\mathrm{i}}\xspace}

 \def\Ps      {\ensuremath{\mathrm{s}}\xspace}

 \def\thebaroffset{0.0em}
}
{
 
 \def\Pgamma      {\ensuremath{\gamma}\xspace}

 \def\Ppi         {\ensuremath{\pi}\xspace}                 
                  
 \def\Prho        {\ensuremath{\rho}\xspace}

 \mathchardef\PDelta="7101
 \mathchardef\PXi="7104
 \mathchardef\PLambda="7103
 \mathchardef\PSigma="7106
 \mathchardef\POmega="710A
 \mathchardef\PUpsilon="7107
 \mathchardef\PPi="7105
                  
 \def\PB      {\ensuremath{B}\xspace}                 
                  
 \def\PD      {\ensuremath{D}\xspace}

 \def\PK      {\ensuremath{K}\xspace}

 \def\Pb      {\ensuremath{b}\xspace}                 
 \def\Pc      {\ensuremath{c}\xspace}

 \def\Ph      {\ensuremath{h}\xspace}                 
 \def\Pi      {\ensuremath{i}\xspace}

 \def\Ps      {\ensuremath{s}\xspace}

 \def\thebaroffset{0.18em}
}
\newcommand{\offsetoverline}[2][\thebaroffset]{\kern #1\overline{\kern -#1 #2}}%

\makeatletter
\ifcase \@ptsize \relax
  \newcommand{\miniscule}{\@setfontsize\miniscule{4}{5}}
\or
  \newcommand{\miniscule}{\@setfontsize\miniscule{5}{6}}
\or
  \newcommand{\miniscule}{\@setfontsize\miniscule{5}{6}}
\fi
\makeatother

\DeclareRobustCommand{\optbar}[1]{\shortstack{{\miniscule (\rule[.5ex]{1.25em}{.18mm})}
  \\ [-.7ex] $#1$}}

\def\g      {{\ensuremath{\Pgamma}}\xspace}

\def\squark    {{\ensuremath{\Ps}}\xspace}

\def\cquark    {{\ensuremath{\Pc}}\xspace}

\def\bquark    {{\ensuremath{\Pb}}\xspace}

\def\hadron {{\ensuremath{\Ph}}\xspace}
\def\pion   {{\ensuremath{\Ppi}}\xspace}

\def\pip    {{\ensuremath{\pion^+}}\xspace}
\def\pim    {{\ensuremath{\pion^-}}\xspace}
\def\pipm   {{\ensuremath{\pion^\pm}}\xspace}
\def\pimp   {{\ensuremath{\pion^\mp}}\xspace}

\def\rhomeson {{\ensuremath{\Prho}}\xspace}

\def\rhopm    {{\ensuremath{\rhomeson^\pm}}\xspace}

\def\kaon    {{\ensuremath{\PK}}\xspace}
\def\Kbar    {{\ensuremath{\offsetoverline{\PK}}}\xspace}

\def\KorKbar {\kern \thebaroffset\optbar{\kern -\thebaroffset \PK}{}\xspace}

\def\Kp      {{\ensuremath{\kaon^+}}\xspace}
\def\Km      {{\ensuremath{\kaon^-}}\xspace}
\def\Kpm     {{\ensuremath{\kaon^\pm}}\xspace}
\def\Kmp     {{\ensuremath{\kaon^\mp}}\xspace}
\def\KS      {{\ensuremath{\kaon^0_{\mathrm{S}}}}\xspace}

\def\Kstarz  {{\ensuremath{\kaon^{*0}}}\xspace}

\def\Kstar   {{\ensuremath{\kaon^*}}\xspace}
\def\Kstarb  {{\ensuremath{\Kbar{}^*}}\xspace}
\def\Kstarp  {{\ensuremath{\kaon^{*+}}}\xspace}
\def\Kstarm  {{\ensuremath{\kaon^{*-}}}\xspace}
\def\Kstarpm {{\ensuremath{\kaon^{*\pm}}}\xspace}

\def\Dbar    {{\ensuremath{\offsetoverline{\PD}}}\xspace}
\def\D       {{\ensuremath{\PD}}\xspace}

\def\DorDbar {\kern \thebaroffset\optbar{\kern -\thebaroffset \PD}\xspace}
\def\Dz      {{\ensuremath{\D^0}}\xspace}
\def\Dzb     {{\ensuremath{\Dbar{}^0}}\xspace}
\def\Dp      {{\ensuremath{\D^+}}\xspace}
\def\Dm      {{\ensuremath{\D^-}}\xspace}

\def\DpDm    {\ensuremath{\Dp {\kern -0.16em \Dm}}\xspace}
\def\Dstar   {{\ensuremath{\D^*}}\xspace}

\def\Dstarz  {{\ensuremath{\D^{*0}}}\xspace}

\def\B       {{\ensuremath{\PB}}\xspace}

\def\BorBbar {\kern \thebaroffset\optbar{\kern -\thebaroffset \PB}\xspace}
\def\Bz      {{\ensuremath{\B^0}}\xspace}

\def\Bd      {{\ensuremath{\B^0}}\xspace}

\def\BdorBdbar {\kern \thebaroffset\optbar{\kern -\thebaroffset \Bd}\xspace}
\def\Bu      {{\ensuremath{\B^+}}\xspace}
\def\Bub     {{\ensuremath{\B^-}}\xspace}
\def\Bp      {{\ensuremath{\Bu}}\xspace}
\def\Bm      {{\ensuremath{\Bub}}\xspace}
\def\Bpm     {{\ensuremath{\B^\pm}}\xspace}

\def\Bs      {{\ensuremath{\B^0_\squark}}\xspace}

\def\BsorBsbar {\kern \thebaroffset\optbar{\kern -\thebaroffset \Bs}\xspace}

\def\Y#1S{\ensuremath{\PUpsilon{(#1S)}}\xspace}

\def\Lz          {{\ensuremath{\PLambda}}\xspace}

\def\LorLbar     {\kern \thebaroffset\optbar{\kern -\thebaroffset \PLambda}\xspace}

\def\Lc          {{\ensuremath{\Lz^+_\cquark}}\xspace}

\def\Lb           {{\ensuremath{\Lz^0_\bquark}}\xspace}

\newcommand{\decay}[2]{\ensuremath{#1\!\to #2}\xspace} 

\def\to                 {\ensuremath{\rightarrow}\xspace}

\def\CP                {{\ensuremath{C\!P}}\xspace}

\def\AT#1     {\ensuremath{A_{\mathrm{T}}^{#1}}\xspace}           

\def\C#1      {\ensuremath{\mathcal{C}_{#1}}\xspace}                       
\def\Cp#1     {\ensuremath{\mathcal{C}_{#1}^{'}}\xspace}                    
\def\Ceff#1   {\ensuremath{\mathcal{C}_{#1}^{\mathrm{(eff)}}}\xspace}        
\def\Cpeff#1  {\ensuremath{\mathcal{C}_{#1}^{'\mathrm{(eff)}}}\xspace}       
\def\Ope#1    {\ensuremath{\mathcal{O}_{#1}}\xspace}                       
\def\Opep#1   {\ensuremath{\mathcal{O}_{#1}^{'}}\xspace}                    

\newcommand{\nospaceunit}[1]{\ensuremath{\text{#1}}}       
\newcommand{\aunit}[1]{\ensuremath{\text{\,#1}}}       

\newcommand{\tev}{\aunit{Te\kern -0.1em V}\xspace}
\newcommand{\gev}{\aunit{Ge\kern -0.1em V}\xspace}
\newcommand{\mev}{\aunit{Me\kern -0.1em V}\xspace}
\newcommand{\kev}{\aunit{ke\kern -0.1em V}\xspace}
\newcommand{\ev}{\aunit{e\kern -0.1em V}\xspace}
 
\newcommand{\mevc}{\ensuremath{\aunit{Me\kern -0.1em V\!/}c}\xspace}
\newcommand{\gevc}{\ensuremath{\aunit{Ge\kern -0.1em V\!/}c}\xspace}
\newcommand{\mevcc}{\ensuremath{\aunit{Me\kern -0.1em V\!/}c^2}\xspace}
\newcommand{\gevcc}{\ensuremath{\aunit{Ge\kern -0.1em V\!/}c^2}\xspace}

\def\mum  {\ensuremath{\,\upmu\nospaceunit{m}}\xspace}

\def\fb   {\ensuremath{\aunit{fb}}\xspace}
\def\invfb   {\ensuremath{\fb^{-1}}\xspace}

\newcommand{\chisq}{\ensuremath{\chi^2}\xspace}

\newcommand{\chisqip}{\ensuremath{\chi^2_{\text{IP}}}\xspace}

\def\gsim{{~\raise.15em\hbox{$>$}\kern-.85em
          \lower.35em\hbox{$\sim$}~}\xspace}
\def\lsim{{~\raise.15em\hbox{$<$}\kern-.85em
          \lower.35em\hbox{$\sim$}~}\xspace}

\def\pt         {\ensuremath{p_{\mathrm{T}}}\xspace}

\def\ptot       {\ensuremath{p}\xspace}

\def\evtgen     {\mbox{\textsc{EvtGen}}\xspace}

\def\geant      {\mbox{\textsc{Geant4}}\xspace}

\def\photos     {\mbox{\textsc{Photos}}\xspace}

\def\pythia     {\mbox{\textsc{Pythia}}\xspace}

\def\tell1  {TELL1\xspace}
\def\ukl1   {UKL1\xspace}

\newcommand{\eg}{\mbox{\itshape e.g.}\xspace}

\newcommand{\lhcborcid}[1]{\href{https://orcid.org/#1}{\hspace*{0.1em}\raisebox{-0.45ex}{\includegraphics[width=1em]{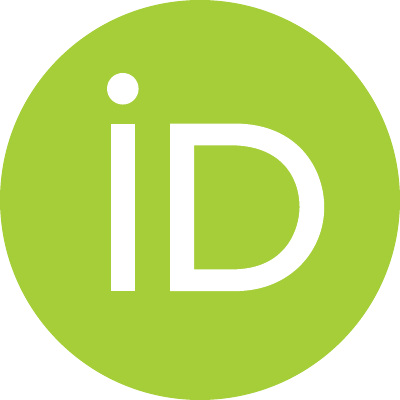}}}}

\usepackage{cite} 
\usepackage{mciteplus}

\usepackage{longtable} 

\usepackage{lscape}
\usepackage{placeins}
\usepackage{siunitx}

\begin{document}

\renewcommand{\thefootnote}{\fnsymbol{footnote}}
\setcounter{footnote}{1}


\begin{titlepage}
\pagenumbering{roman}

\vspace*{-1.5cm}
\centerline{\large EUROPEAN ORGANIZATION FOR NUCLEAR RESEARCH (CERN)}
\vspace*{1.5cm}
\noindent
\begin{tabular*}{\linewidth}{lc@{\extracolsep{\fill}}r@{\extracolsep{0pt}}}
\ifthenelse{\boolean{pdflatex}}
{\vspace*{-1.5cm}\mbox{\!\!\!\includegraphics[width=.14\textwidth]{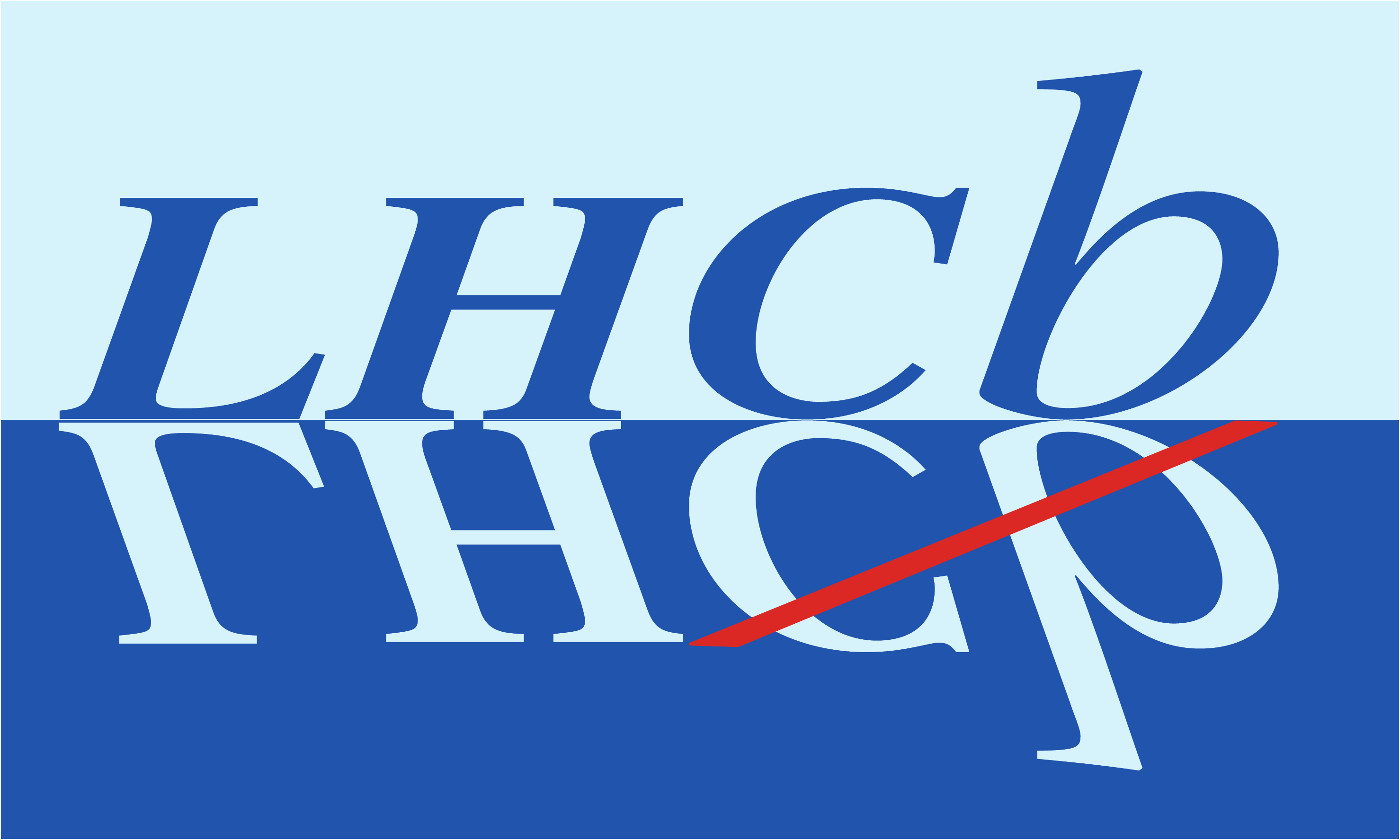}} & &}%
{\vspace*{-1.2cm}\mbox{\!\!\!\includegraphics[width=.12\textwidth]{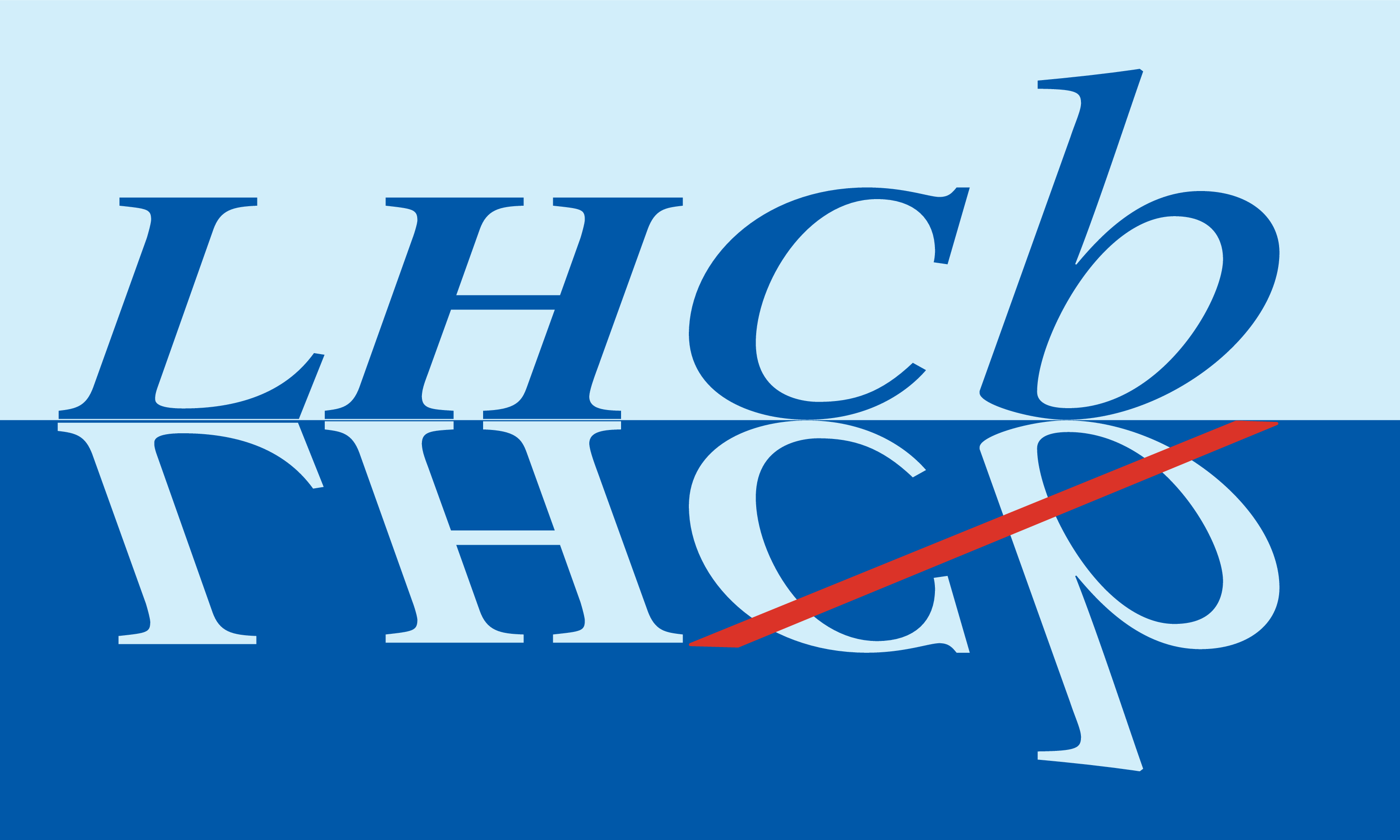}} & &}%
\\
 & & CERN-EP-2024-260 \\  
 & &  LHCb-PAPER-2024-023 \\  
 & & February 18, 2025 \\ 
 & & \\
\end{tabular*}

\vspace*{4.0cm}

{\normalfont\bfseries\boldmath\huge
\begin{center}
  \papertitle 
\end{center}
}

\vspace*{2.0cm}

\begin{center}
\paperauthors\footnote{Authors are listed at the end of this paper.}
\end{center}

\vspace{\fill}

\begin{center}
    This paper is dedicated to the memory of our friend and colleague, \\ Dr. Arnau Brossa Gonzalo, whose promising career was tragically cut short.
\end{center}

\vspace{\fill}

\begin{abstract}
  \noindent
  Measurements of ${\ensuremath{C\!P}}$ observables and the CKM angle $\gamma$ are performed in \mbox{$B^{\pm} \to D K^*(892)^{\pm}$} decays, where $D$ represents a superposition of $D^0$ and ${\ensuremath{\offsetoverline{D}}{}^0}$ states, using the \mbox{LHCb} dataset collected during Run 1 (2011\textendash2012) and Run 2 (2015\textendash2018). A study of this channel is presented with the $D$ meson reconstructed in two-body final states $K^{\pm}\pi^{\mp}$, $K^+K^-$ and $\pi^+\pi^-$; four-body final states $K^{\pm}\pi^{\mp}\pi^{\pm}\pi^{\mp}$ and $\pi^+\pi^-\pi^+\pi^-$; and three-body final states $K^0_{\mathrm{S}} \pi^+\pi^-$ and $K^0_{\mathrm{S}} K^+ K^-$. This analysis includes the first observation of the suppressed \mbox{$B^{\pm} \to [\pi^{\pm}K^{\mp}]_D K^{*\pm}$} and \mbox{$B^{\pm} \to [\pi^{\pm}K^{\mp}\pi^{\pm}\pi^{\mp}]_D K^{*\pm}$} decays. The combined result gives $\gamma=(63\pm 13)^\circ$.
  
\end{abstract}

\vspace*{2.0cm}

\begin{center}
  Published in JHEP 02 (2025) 113 
\end{center}

\vspace{\fill}

{\footnotesize 
\centerline{\copyright~\papercopyright. \href{\paperlicenceurl}{\paperlicence}.}}
\vspace*{2mm}

\end{titlepage}


\newpage
\setcounter{page}{2}
\mbox{~}
%
%
%
%


\renewcommand{\thefootnote}{\arabic{footnote}}
\setcounter{footnote}{0}

\cleardoublepage


\pagestyle{plain} 
\setcounter{page}{1}
\pagenumbering{arabic}


\def\btodk {\ensuremath{\decay{\Bpm}{\D \Kpm}}\xspace}
\def\btodpi {\ensuremath{\decay{\Bpm}{\D \pipm}}\xspace}
\def\btodh {\ensuremath{\decay{\Bpm}{\D \hadron^{\pm}}}\xspace}
\def\btodkstar {\ensuremath{\decay{\Bpm}{\D \Kstarpm}}\xspace}
\def\btodstkstar {\ensuremath{\decay{\Bpm}{\Dstar \Kstarpm}}\xspace}
\def\btodkstarz {\ensuremath{\decay{\Bz}{\D \Kstarz}}\xspace}
\def\btodstarh {\ensuremath{\decay{\Bpm}{\Dstar \hadron^{\pm}}}\xspace}

\section{Introduction}
\label{sec:Introduction}
 A precise determination of the Cabibbo--Kobayashi--Maskawa (CKM)~\cite{Cabibbo:1963yz,Kobayashi:1973fv} angle $\gamma$ is essential to test the description of $\CP$ violation in the Standard Model~(SM). This SM benchmark parameter is determined in the interference between processes involving the transitions \mbox{$\decay{b}{c \bar{u}s}$} and \mbox{$\decay{b} {u\bar{c}s}$}. The decay channels most sensitive to these interference effects are \mbox{$\btodk$} decays, where the $\D$ meson is a superposition of $\Dz$ and $\Dzb$ mesons that decay to the same final state. These tree-level decays provide a direct measurement of $\gamma$ that is unlikely to be altered by the presence of physics beyond the SM. Direct measurements can be compared to an indirect determination of $\gamma$, which is calculated from the assumption of unitarity and measurements of other parameters in the CKM matrix. 

 A recent \lhcb combination of direct measurements is $\g = (63.8_{-3.7}^{+3.5})^{\circ}$~\cite{LHCb-CONF-2022-003}. Global fits performed by the CKMfitter and UTfit groups, where all the direct determinations are excluded, return the value $\g = (66.29_{-1.86}^{+0.72})^{\circ}$~\cite{CKMfitter2015} and $\g = (64.9 \pm 1.4)^{\circ}$~\cite{UTfit2023}, respectively. 
 The current uncertainty on the direct measurement remains larger than that of the indirect determination. This is due to the small branching fraction of decays sensitive to $\gamma$. So far, the focus has been on \mbox{$\btodk$} and \mbox{$\btodkstarz$} decays~\cite{LHCb-PAPER-2020-019, LHCb-PAPER-2022-017, LHCb-PAPER-2022-037, LHCb-PAPER-2023-009, LHCb-PAPER-2023-012, LHCb-PAPER-2023-029, LHCb-PAPER-2023-040} which drive the precision. This paper presents a measurement of $\gamma$ using \mbox{$\btodkstar$} decays, where $\Kstarpm$ refers to the $K^*(892)^{\pm}$ resonance. This decay has a similar branching fraction and sensitivity to $\gamma$ as \mbox{$\btodk$} but is less well studied. 
 At the \lhcb experiment, the observed yields of \mbox{\btodkstar} decays are lower than those of \mbox{\btodk} due to the differences in reconstruction efficiency between a $\Kstarpm$ and $\Kpm$ meson. The former is reconstructed through the decay \mbox{$\decay{\Kstarpm}{\KS(\to \pip \pim)\pipm}$} which requires the reconstruction of an additional $\KS$ meson. Nonetheless, the extremely low level of backgrounds makes this decay attractive for further study. 
 It has been previously analysed by the $B$ factories with two- and three-body $D$ final states. The Belle collaboration has studied the decay \mbox{$D \to \KS\pip\pim$}~\cite{belle_prev}, whereas the BaBar collaboration has analysed a range of $D$ decays~\cite{babar_prev,BaBar_adsglw_prev}. A set of two- and four-body $D$ decays have already been investigated at \lhcb~\cite{LHCb-PAPER-2017-030} with a dataset corresponding to an integrated luminosity of 4.8\invfb at centre-of-mass energies $\sqrt{s}=$ 7, 8 and 13\tev.
 
 This paper makes use of a mixture of $\D$-decay final states that can be grouped into three categories: (quasi-)$\CP$ eigenstates \mbox{$\Kp\Km$}, \mbox{$\pip\pim$} and (\mbox{$\pip\pim\pip\pim$}), the self-conjugate states \mbox{$\KS h^+ h^-$}, where $h=\pi, K$, and states that contain charged kaons and pions and are not self-conjugate, \mbox{$\Km \pip$}, \mbox{$\pim\Kp$}, \mbox{$\Km \pip \pim \pip$} and \mbox{$\pim\Kp\pim\pip$}. Quasi-$\CP$ eigenstates refer to $\D$-decay modes which are not pure $\CP$ eigenstates, but can be treated as though they were as long as the $\CP$-even fraction is known. For the non-self-conjugate $\D$-decay modes, where the charge of the kaon is the same as the $\B$-meson charge, the decay modes are referred to as same sign (SS) and denoted as \mbox{$D \to K \pi$} or \mbox{$D \to K \pi\pi\pi$}. When the kaon and $\B$-meson charges are opposite, the decay modes are referred to as opposite sign (OS) and denoted as \mbox{$D \to \pi K$} or \mbox{$D \to \pi K \pi\pi$}. The OS decay modes are suppressed with respect to SS modes. This measurement uses the \lhcb dataset of proton-proton ($pp$) collisions collected during Run~1 (2011\textendash2012) and Run~2 (2015\textendash2018), corresponding to an integrated luminosity of 9\invfb at centre-of-mass energies $\sqrt{s}=$ 7, 8 and 13\tev. The $\CP$-violation-sensitive observables are measured independently for each decay channel. Then they are combined to obtain the value of the angle $\g$. 
 
 The measurement strategies for the different decay channels are given in Sec.~\ref{sec:method}. The \lhcb detector is described in Sec.~\ref{sec:detector} and the selection of \mbox{$\btodkstar$} candidates is detailed in Sec.~\ref{sec:selection}. Section~\ref{sec:bkg} elaborates the backgrounds that are present after selection followed by the details of signal yields for each mode in Sec.~\ref{sec:yields}. The determination of $\CP$-violating observables is described in Sec.~\ref{sec:cpfit}. Systematic uncertainties are discussed in Sec.~\ref{sec:syst}. The results and their interpretation are shown in Sec.~\ref{sec:results}.

\section{Analysis overview}
\label{sec:method}

The amplitude of a $\Bm$ decay to $\Dz \Kstarm$ decay is defined as
\begin{equation}
\mathcal{A}(\Bm \to \Dz \Kstarm) \equiv A_B,
\end{equation}
and that of a $\Bm$ decay to $\Dzb \Kstarm$ decay is given as
\begin{equation}
\mathcal{A}(\Bm \to \Dzb \Kstarm) = A_B \cdot r_B \cdot e^{i(\delta_B - \gamma)},
\end{equation}

\noindent
where $r_B$ is the ratio of the magnitudes of the amplitudes of \mbox{$\Bm \to \Dzb \Kstarm$} and \mbox{$\Bm \to \Dz \Kstarm$} and $\delta_B$ is the corresponding strong-phase difference between them. When the $\D$ meson decays to a state that is accessible from both the $\Dz$ and $\Dzb$ meson,  there is interference between the two decay paths. This interference gives access to the CKM angle $\gamma$. Analogous expressions can be formed for the charge-conjugate $\Bp$ decay where the sign of the \g angle is flipped ($\g \leftrightarrow - \g$). The measurement strategies for the various types of $D$ decays are described in detail in this section. 

The amplitudes for the $\Dz$ and the $\Dzb$ mesons decaying to $\CP$ eigenstates are the same.\footnote{Violation of \CP symmetry in $D$ decays is neglected. This effect is much smaller compared to \CP violation in the $B$-meson decay.} Hence, the difference in the two overall amplitudes is driven by the difference in the two $B$-meson amplitudes. This leads to a relatively small amount of $\CP$ violation that is of the order of $r_B$. 
A coherence factor $\kappa$ is defined, satisfying $0 \le \kappa \le 1$, which quantifies the dilution of the interference effects of interest due to selected \mbox{$\Bpm \to D \KS \pipm$} decays that do not proceed through the intermediate $K^*(892)^{\pm}$ resonance. The value of $\kappa$ is estimated in Ref.~\cite{LHCb-PAPER-2017-030} to be $0.95\pm0.06$. A coherence value of one would represent a pure resonant sample. 

Measurements with \emph{CP} eigenstates are performed according to the formalism presented in Refs.~\cite{GLW_ref1,GLW_ref2}. This includes obtaining the asymmetries between the rates of $\Bp$ and $\Bm$ decays and ratios with SS decays that are directly related to the physics parameters of interest. The technique can be extended to quasi-\CP eigenstates provided the \CP-even fraction of the final state, $F_{\CP}$, is known~\cite{qGLW}. For \mbox{\btodkstar}, when the $D$ meson decays to a final state $X$, the ratio $R_{\CP}^X$ is defined as
\begin{equation}
\begin{split}
    R_{\CP}^X &= \frac{\Gamma(\Bm \to [X]_D \Kstarm) + \Gamma(\Bp \to [X]_D \Kstarp)}{\Gamma(\Bm \to [SS]_D \Kstarm) + \Gamma(\Bp \to [SS]_D \Kstarp)} \times \frac{\mathcal{B}(\Dz \to SS)}{\mathcal{B}(\Dz \to X)} \\ &= 1 + r_{B}^{2} + 2 \kappa (2F_{\CP}^X-1)r_{B} \cos{\delta_{B}} \cos{\gamma},
\end{split}
\label{eq:R_kk}
\end{equation}
and the asymmetry $A_{\CP}^X$ as
\begin{equation}
\begin{split}
    A_{\CP}^X &= \frac{\Gamma(\Bm \to [X]_D \Kstarm) - \Gamma(\Bp \to [X]_D \Kstarp)}{\Gamma(\Bm \to [X]_D \Kstarm) + \Gamma(\Bp \to [X]_D \Kstarp)}\\ &= \frac{2 \kappa (2F_{\CP}^X -1)r_{B} \sin{\delta_{B}} \sin{\gamma}}{R_{\CP}^X}.
\end{split}
\label{eq:A_kk}
\end{equation}
For \mbox{$\D \to KK$} and \mbox{$\D \to \pi\pi$}, the \CP-even fraction is one. For \mbox{$\D \to \pi\pi\pi\pi$}, the \CP-even fraction has been measured at the \besiii experiment as $0.735 \pm 0.016$~\cite{BESIIIFPlus4Pi}. The ratio $R_{\CP}^X$ and asymmetry $A_{\CP}^X$ are measured for each $\D$-meson decay mode that is a (quasi)-\CP eigenstate, and the SS decay chosen for the normalisation is the one with the same multiplicity as the (quasi-)\CP eigenstate. 

The formalism for measuring $\gamma$ using the OS and SS $D$-meson decays is given in Refs.\mbox{~\cite{ADS_ref1,ADS_ref2}}. For the OS decay channels, the two amplitudes from the \B decay to the final state are of similar size, resulting in maximum interference.  In this formalism, the hadronic properties of the $D$ decay become relevant to the interpretation of the ratios with SS decays and asymmetries between \Bp and \Bm decays. The hadronic parameters are defined as $r_D^X$, $\delta_D^X$, and $\kappa^X$, where $X$ is the final state, $r_D^X$ is the ratio of the suppressed and favoured $D$-decay amplitudes, $\delta_D^X$ is the average strong-phase difference between them and $\kappa_D^X$ takes into account the variation of $\delta_D^X$ over the phase space which dilutes the interference. Since \mbox{$D\to K\pi$} is a two-body decay, the value of $\kappa^X$ is unity. The ratios and asymmetries between the OS and SS decay modes are given by

\begin{equation}
\begin{split}
    R_{OS}^X &= \frac{\Gamma(\Bm \to [OS]_D \Kstarm) + \Gamma(\Bp \to [OS]_D \Kstarp)}{\Gamma(\Bm \to [SS]_D \Kstarm) + \Gamma(\Bp \to [SS]_D \Kstarp)} \\ &= r_{B}^{2} + (r_{D}^{X})^{2} + 2 \kappa r_{B} \kappa_D^{X} r_{D}^{X} \cos{(\delta_{B} + \delta_{D}^{X})} \cos{\gamma},
\end{split}
\label{eq:R_ads}
\end{equation}

\begin{equation}
\begin{split}
    A_{OS}^X &= \frac{\Gamma(\Bm \to [OS]_D \Kstarm) - \Gamma(\Bp \to [OS]_D \Kstarp)}{\Gamma(\Bm \to [OS]_D \Kstarm) + \Gamma(\Bp \to [OS]_D \Kstarp)} \\ &= \frac{2 \kappa r_{B} \kappa_D^{X}r_{D}^{X} \sin{(\delta_{B} + \delta_{D}^{X})} \sin{\gamma}}{R_{OS}^X},
 \end{split}   
\label{eq:A_ads}
\end{equation}

\begin{equation}
\begin{split}
    A_{SS}^X &= \frac{\Gamma(\Bm \to [SS]_D \Kstarm) - \Gamma(\Bp \to [SS]_D \Kstarp)}{\Gamma(\Bm \to [SS]_D \Kstarm) + \Gamma(\Bp \to [SS]_D \Kstarp)} \\ &= \frac{2 \kappa r_{B} \kappa_D^{X} r_{D}^{X} \sin{(\delta_{B} - \delta_{D}^{X})} \sin{\gamma}}{1 + r_{B}^{2} (r_{D}^{X})^{2} + 2 \kappa r_{B} \kappa_D^{X} r_{D}^{X} \cos{(\delta_{B} - \delta_{D}^{X})} \cos{\gamma}}. 
\end{split}
\label{eq:A_adsfav}
\end{equation}
Taking expected values for $r_B$ and $r_D$~\cite{CKMfitter2015}, it is clear that $A_{SS}^X$ is small in comparison to $A_{OS}^X $. Hence, the sensitivity to \CP violation in the SS decay modes is expected to be very small.

For the three-body decay modes, the variance of the strong phase $\delta_D$ over the phase space of the decay is large. This leads to values of $F_{\CP}$ close to 0.5 which maximally dilutes the interference term, and hence, measuring $R_{CP}^{\KS h^+h^-}$ and $A_{CP}^{\KS h^+h^-}$ will result in very little sensitivity to $\gamma$. Instead, these decays are analysed in regions of phase space to exploit regions where the coherence is high. The formalism is developed in Refs.~\cite{BPGGSZ_ref1,BPGGSZ_ref2,BPGGSZ_ref3}. The final state is dependent on the kinematics of the decay products which can be visualised using the Dalitz plot. The strong phase varies over the phase space, and the strong-phase difference at a point in phase space between the \Dz and \Dzb decay is given by $\Delta{\delta_D}$. The density of the $B^-$ decay, $P_{B}$, at a specific point in the $D$ phase space is expressed as
\begin{equation}
\begin{split}
P_{B}  &=  |{A}|^{2} + r_{B}^{2}|\overline{A}|^{2} + r_{B}({A}^{*}\overline{A}e^{i(\delta_{B}-\gamma)} + {A}\overline{A}^{*}e^{-i(\delta_{B}-\gamma)}) \\  &= |{A}|^{2} + r_{B}^{2}|\overline{A}|^{2} + 2\kappa r_{B}|\overline{A}||A|(\cos\Delta \delta_{D} \cos(\delta_{B}-\gamma) - \sin\Delta \delta_{D} \sin(\delta_{B}-\gamma)) \\
 &= |{A}|^{2} + r_{B}^{2}|\overline{A}|^{2} + 2\kappa \sqrt{|A|^2||\overline{A}|^2}(C \, x_{-} - S\,y_{-}),
 \end{split}
\label{Eq:pB2}
\end{equation}
where $A$ and $\overline{A}$ are the amplitudes of the $\Dz$ and $\Dzb$ decays to the \mbox{$\KS h^+h^-$} final state, \mbox{$x_{-} = r_{B}\cos(\delta_{B}-\gamma)$}, \mbox{$y_{-} = r_B\sin(\delta_{B}-\gamma)$}, \mbox{$C = \cos\Delta \delta_{D}$} and \mbox{$S = \sin\Delta \delta_{D}$}. For the charge-conjugate mode, \mbox{$\Bu \to \D\Kstarp$}, the density is given by the same expression with \mbox{$A \leftrightarrow \overline{A}$}, \mbox{$\gamma \leftrightarrow -\gamma$}, \mbox{$x_- \leftrightarrow x_+$} and \mbox{$y_- \leftrightarrow y_+$}, where \mbox{$x_{+} = r_{B}\cos(\delta_{B}+\gamma)$} and \mbox{$y_{+} = r_B\sin(\delta_{B}+\gamma)$}.

The $D$-meson decay Dalitz plot is split into bin pairs, symmetric with respect to the \mbox{$m^2_{\KS h^+}=m^2_{\KS h^-}$} axis. Bins with \mbox{$m^2_{\KS h^-} > m^2_{\KS h^+}$} are given positive indices 1 to $N$. Those with \mbox{$m^2_{\KS h^-} < m^2_{\KS h^+}$} are given negative indices such that bins $i$ and $-i$ are \CP conjugates. The partial decay rate in the $i^{\text{th}}$ bin of the $D$-decay phase space for $\Bm$ and $\Bu$ modes, $\Gamma_i^\mp$, can be found by integrating Eq.~\ref{Eq:pB2} and written as
\begin{equation}
\Gamma_{i}^{\mp} 
\propto  \left(F_{\pm i} + (r_{B})^{2}F_{\mp i} +2\kappa\sqrt{F_{i}F_{-i}}(c_{i}x_{\mp} \pm s_{i}y_{\mp})\right).  \label{Eq:B-}
\end{equation}
Here, $F_{\pm i}$ is the fraction of observed $\Dz$-meson decays in each bin given the experimental reconstruction and selection efficiency, and migration effects. Neglecting $\CP$ violation in these charm decays, the charge conjugate decays satisfy $F_{i} = \overline{F}_{-i}$, where $\overline{F}_{i}$ is the fraction of observed $\Dzb$ decays in the bin with index $i$. The $c_i$ and $s_i$ parameters are the amplitude-weighted averages of the functions $C$ and $S$ over the bin. 

The values of $c_i$ and $s_i$ have been measured by the \cleo~\cite{CLEO} and \besiii~\cite{BESIII,bes3kk} collaborations. Both have followed the same partition of the phase space. There are multiple options presented in Ref.~\cite{CLEO}. The ones used in this analysis are named the ``optimal'' and \mbox{``2-bins''} binning schemes for \mbox{$D \to \KS \pip \pim$} and \mbox{$\D \to \KS \Kp \Km$} decays, respectively. 
These represent the schemes that will result in the best sensitivity for $\gamma$ under the experimental conditions for this analysis. There are in total \mbox{$2\times 8 = 16$} and \mbox{$2\times 2 =4$} bins in the \mbox{$D \to \KS\pip\pim$} and \mbox{$D \to \KS\Kp\Km$} modes, respectively. 
These schemes are followed in this analysis so that the values $c_i$ and $s_i$ can be directly used as external inputs. The $F_{\pm i}$ values are taken from the \mbox{$\Bpm \to D(\to \KS h^{\pm}h^{\mp})h^{\pm}$} analysis~\cite{LHCb-PAPER-2020-019}. 
This assumes that the relative efficiency between two points on the Dalitz plot of the $D$ decay is the same for \mbox{$\Bpm \to Dh^{\pm}$} and \mbox{$\Bpm \to D\Kstarpm$} modes, which is verified in simulation. The use of these external inputs, rather than reliance on a model of the $D$-meson decay, ensures that the measurement is amplitude-model independent and does not incur systematic uncertainties due to the model which are hard to quantify.

The SS decay modes are used as a control channel in this analysis. Their high yield benefits the reconstructed-mass parametrisation, which is used to model the small background contributions from \mbox{\btodstkstar} decays in the other $D$-decay modes. The \CP observables are measured in the \CP fit, consisting of a simultaneous maximum-likelihood fit performed in the different categories of the two- and four-body decays defined by the $B$ charge and $D$-decay mode. The ratios and asymmetries in Eqs.\mbox{~\ref{eq:R_kk}\textendash\ref{eq:A_adsfav}} are measured. Another fit is performed simultaneously for the three-body decays in the categories of $B$ charge, $D$-decay mode and Dalitz plot bin. The $\CP$ observables $x_{\pm}$ and $y_{\pm}$ are extracted from this fit. The results from all sets of modes are combined to obtain a precise single solution for $\gamma$.
 
\section{Detector and simulation}
\label{sec:detector}

The \lhcb detector~\cite{LHCb-DP-2008-001,LHCb-DP-2014-002} is a single-arm forward
spectrometer covering the \mbox{pseudorapidity} range $2<\eta <5$,
designed for the study of particles containing \bquark or \cquark
quarks. The detector includes a high-precision tracking system
consisting of a silicon-strip vertex detector surrounding the $pp$
interaction region~\cite{LHCb-DP-2014-001}, a large-area silicon-strip detector located
upstream of a dipole magnet with a bending power of about
$4{\mathrm{\,T\,m}}$, and three stations of silicon-strip detectors and straw
drift tubes~\cite{LHCb-DP-2013-003,LHCb-DP-2017-001}
placed downstream of the magnet.
The tracking system provides a measurement of the momentum, \ptot, of charged particles with
a relative uncertainty that varies from 0.5\% at low momentum to 1.0\% at 200\gevc.
The minimum distance of a track to a primary $pp$ collision vertex (PV), the impact parameter (IP), 
is measured with a resolution of $(15+29/\pt)\mum$,
where \pt is the component of the momentum transverse to the beam, in\,\gevc.
Different types of charged hadrons are distinguished using information
from two ring-imaging Cherenkov detectors~\cite{LHCb-DP-2012-003}.
Photons, electrons and hadrons are identified by a calorimeter system consisting of
scintillating-pad and preshower detectors, an electromagnetic
and a hadronic calorimeter. Muons are identified by a
system composed of alternating layers of iron and multiwire
proportional chambers~\cite{LHCb-DP-2012-002}.

The online event selection is performed by a trigger~\cite{LHCb-DP-2012-004}, which consists of a hardware stage, based on information from the calorimeter and muon
systems, followed by a software stage, which applies a full event
reconstruction. The events for the analysis that satisfy the hardware trigger are selected through two routes. Either they have an energy deposit in the calorimeter associated to the signal decay, or a particle not associated to the signal candidate fulfills any requirement. At the software stage, at least one particle is required to have high $p_T$ and high \chisqip, where \chisqip is defined as the difference in the primary vertex fit \chisq with and without the inclusion of that particle. 
A multivariate algorithm~\cite{BBDT} is used to select secondary vertices consistent with being a two-, three-, or four-track $\bquark$-hadron decay.

 Simulation is required to model the effects of the detector acceptance and the imposed selection requirements.
  In the simulation, $pp$ collisions are generated using
  \pythia~\cite{Sjostrand:2007gs,*Sjostrand:2006za}  with a specific \lhcb configuration~\cite{LHCb-PROC-2010-056}.
  Decays of unstable particles
  are described by \evtgen~\cite{Lange:2001uf}, in which final-state
  radiation is generated using \photos~\cite{davidson2015photos}.
  The interaction of the generated particles with the detector, and its response,
  are implemented using the \geant
  toolkit~\cite{Allison:2006ve, *Agostinelli:2002hh} as described in
  Ref.~\cite{LHCb-PROC-2011-006}.

\section{Candidate selection}
\label{sec:selection}

All tracks and decay vertices are required to be of good quality, and the reconstructed  mass of the \KS, \D and \Kstarpm candidates must be close to their known values~\cite{PDG2024}.
The \KS candidates are formed from two oppositely charged pion tracks, that are reconstructed either using hits in the vertex detector and other downstream tracking stations, or only the latter. These track types are referred to as \textit{long} and \textit{downstream}, respectively. Requirements are placed on the reconstructed \KS mass to be within $\pm$15(20)\mevcc of the known mass for \textit{long} (\textit{downstream}) candidates. 
The \D-meson candidates are reconstructed in two-, four- and three-body final-state categories, labelled $D \to hh$, $D \to 4h$, and $D \to \KS hh$. 

In the case of \mbox{$D \to hh$} decays, the \D-meson candidate is formed by combining two oppositely charged tracks. Particle identification~(PID) requirements are placed on the kaons and pions in such a way that each pair of tracks can only appear in one \D-meson decay sample. The particle selection does not sufficiently suppress the double misidentification of $\Km\pip$ as $\pim \Kp$, which peaks at the \D-meson mass. Candidates in the OS sample are removed if the measured \D-meson mass, when reconstructed under the double misidentification hypothesis, falls within $\pm$15\mevcc of the known \D-meson mass~\cite{PDG2024}. The corresponding selection is not made in the SS sample since here the contamination rate is very low. 

For the \mbox{$\D \to 4h$ decays}, two positively and two negatively charged tracks are combined. PID requirements are placed on the kaons. In the case of \mbox{$D \to K\pi\pi\pi$}, PID requirements are only imposed on those pion candidates that have the opposite charge of the kaon. Additional PID selection on the other pion would result in lower signal efficiency without any significant reduction in misidentified backgrounds. A mass veto of $\pm$15\mevcc around the known \D-meson mass is placed on the OS samples to reduce doubly misidentified background. In the case of the \mbox{$\D \to \pi\pi\pi\pi$} decay, only the two pions with opposite charge to the parent $B$ meson have PID requirements applied. This is sufficient to reduce misidentified decays to a negligible level. Furthermore, candidates are vetoed if the reconstructed mass of any oppositely charged pair of pions is within the range \mbox{480$\textendash$505\mevcc} to remove background from \mbox{$\D \to \KS \pi\pi$}.

For \mbox{$\D \to \KS hh$} decays the candidate is formed by combining a \KS candidate with two oppositely charged tracks, where additionally the $\KS$ mass is constrained to its known value in a kinematic fit~\cite{DTF}. For the two tracks originating from the \D-meson vertex, PID requirements are placed to reduce background from \mbox{$\D\to\KS\Kp\pim$} decays, semileptonic \D decays, and decays in flight of hadrons to leptons. 
In all cases the reconstructed mass of the \D-meson candidate is required to be within $\pm$25\mevcc of the known mass~\cite{PDG2024}. 

A candidate \Kstarpm meson is reconstructed by combining a \KS and a companion pion. The reconstructed mass is required to be within $\pm$75\mevcc of the known $\Kstarpm$ mass~\cite{PDG2024}. The \D-meson candidate is combined with the \Kstarpm to form the \Bpm-meson candidate. To suppress background from the \mbox{$B\to D\KS K$} decays, PID requirements are placed on the companion pion. Background from nonresonant \mbox{$\Bpm \to \D \KS \pipm$} decays is reduced by using the properties of the vector-particle decay. The helicity angle $\theta$ is defined as the angle between the vectors of the companion \KS meson and the $B$ momentum in the \Kstarpm rest frame. It is required that $\lvert\cos \theta\rvert > 0.3$. 

A requirement is placed on the displacement of the \D-meson vertex from the \mbox{\B-meson} vertex to reduce background from \B decays to the final-state particles without the intermediate \D meson. A displacement requirement from their parent meson is also applied to any \textit{long} \KS meson vertex. This removes background from \mbox{$B \to \D \pi\pi\pi$} decays and suppresses the reconstruction of \mbox{$D\to \pi\pi hh$} as \mbox{$D\to \KS hh$}. Finally, the reconstructed $B$-meson mass is computed applying mass constraints to the \D and the \KS mesons in each decay chain, and constraining the \B meson to originate from the primary vertex. 

A separate boosted decision tree (BDT) classifier~\cite{Breiman,AdaBoost} is optimised for each multiplicity of \D-meson decay. The BDTs are trained using simulation to provide the signal distributions and candidates in the upper sideband of reconstructed $B$ mass in data to provide the background distributions. For \mbox{$\D \to \KS hh$} decays the variables used in the training are the same as those in Ref.~\cite{LHCb-PAPER-2020-019} with the addition of kinematic variables for the companion \KS meson. This is to maintain a similar relative efficiency over the \D-meson phase space, which allows the use of the $F_i$ parameters measured in Ref.~\cite{LHCb-PAPER-2020-019}. 
Similar variables are used for the two- and four-body \D-meson decays. A $k$-fold of two is used to avoid any sculpting of the final upper sideband distribution~\cite{kfolding} since the training sample and fit sample are not mutually exclusive.

Selection requirements on the BDT classifiers are placed based on the optimisation of the sensitivity to the \CP observables. Pseudoexperiments are generated with yields corresponding to different selection values for the output of the BDT classifier and \CP observables are determined in each of them as described in Sec.~\ref{sec:cpfit}. The criterion that gives the best precision on \CP observables is chosen. The resulting BDT selection requirement is different for the OS and (quasi-)\CP eigenstates, the SS decay modes and the $\D \to \KS hh$ decays. The signal selection efficiency is higher than 90\% for each BDT classifier.

\section{Background determination}
\label{sec:bkg}

\begin{figure}[t]
    \centering
    \includegraphics[width=0.75\linewidth]{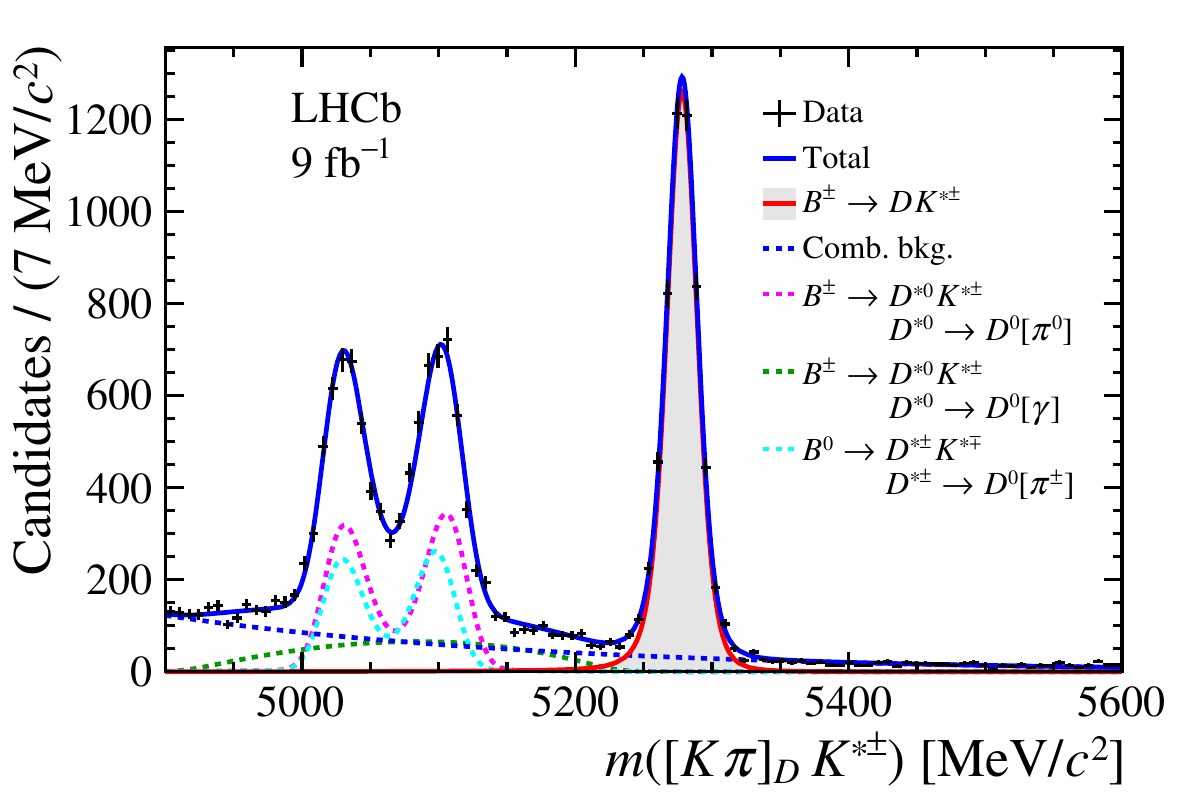}
    \caption{Reconstructed-mass distribution of the two-body favoured (SS) mode $K\pi$ with the fit result superimposed, using the full Run 1 and Run 2 datasets. In the legend, missing particles which are not reconstructed are shown inside square brackets.}
    \label{fig_global_fit_kpi}
\end{figure}

\begin{figure}
    \centering
    \includegraphics[width=0.75\linewidth]{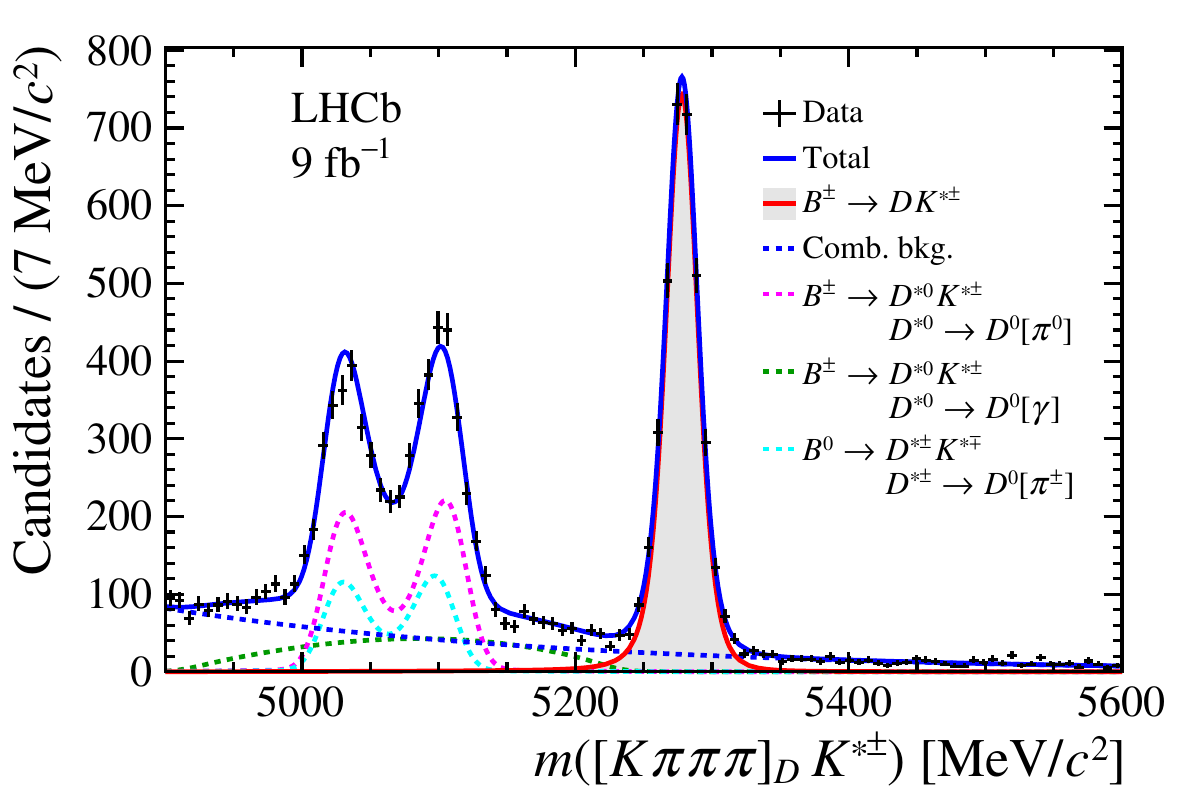}
    \caption{Reconstructed-mass distribution of the four-body favoured (SS) mode $K\pi\pi\pi$ with the fit result superimposed, using the full Run 1 and Run 2 dataset. In the legend, missing particles which are not reconstructed are shown inside square brackets.}
    \label{fig_global_fit_kpipipi}
\end{figure}

For each decay channel, data from the \textit{long} and \textit{downstream} \KS categories are combined. 
Due to the impact of the mass constraints, the effect of the different resolutions of the reconstructed \KS mass in these decays is reduced, and the differences in the reconstructed-mass distributions of signal decays are negligible. A notable background for \mbox{$\btodk$} decays is \mbox{$\btodpi$} with a pion misidentified as a kaon. The corresponding background in \mbox{$\btodkstar$} decays is \mbox{\Bpm \to \D \rhopm}, but since the subsequent decay is different, it does not enter the sample.
The two- and four-body SS decay channels have the highest yields. From the distributions shown in Figs.~\ref{fig_global_fit_kpi} and \ref{fig_global_fit_kpipipi}, it is seen that the signal peak is well separated from \mbox{$B\to\Dstar\Kstar$} decays that are found at lower reconstructed mass due to the missing pion or photon which is not reconstructed. These are called partially reconstructed backgrounds. 
Therefore, the \CP observables are determined using a fit to the reconstructed mass in the range \mbox{5230\textendash5600\mevcc}. 
This removes most of the partially reconstructed background. The combinatorial background level is also low. In order to determine the residual component of the partially reconstructed background, a fit of the SS decay channel is performed over a wider fit range. 
These decays are chosen as they should only exhibit a small rate of \CP violation. 
This is also true for the partially reconstructed decays in the SS decay mode, and therefore their distributions can be fit reliably with \CP observables. 
This is not the case in decay modes with higher rates of \CP violation as there are three amplitudes in the \mbox{$\Bpm \to \Dstarz \Kstarpm$} decays, each with its own strong phase difference. 
Two of these amplitudes have identical reconstructed-mass distributions and hence it is not possible to separate them using this fit, nor to measure the \CP-violating parameters of the \mbox{$\Bpm \to \Dstarz\Kstarpm$} decays. 
Instead, the total amount of partially reconstructed background in the SS candidates is fit, and the amount leaking beyond a reconstructed mass of 5230\mevcc is determined. 
The corresponding background in other decays is estimated by taking this yield and scaling by the corresponding branching fraction and selection efficiency ratios. 

The fit for the reconstructed-mass distribution of SS decay modes is performed as an unbinned extended maximum-likelihood fit. The probability density function (PDF) for the signal shape is given by the sum of a Gaussian and a modified Cruijff function~\cite{LHCb-PAPER-2020-019} defined as
\begin{equation}
f(m) \propto
\begin{cases}
\exp\left[\frac{-\left(m - \mu\right)^2 \left( 1+\beta \left(m - \mu\right)^2 \right)}{2\sigma^2 + \alpha_L \left(m - \mu\right)^2}\right] & \mathrm{if} \, m - \mu < 0,\\[10pt]
\exp\left[\frac{-\left(m - \mu\right)^2 \left( 1+\beta \left(m - \mu\right)^2 \right)}{2\sigma^2 + \alpha_R \left(m - \mu\right)^2}\right] & \mathrm{if} \, m - \mu > 0.\\[10pt]
\end{cases}
\end{equation}

\noindent
The parameters $\mu$ and $\sigma$ are shared between these two components. The tail parameters $\alpha_L$, $\alpha_R$ and $\beta$ of the modified Cruijff function are determined from simulation. In the fit to data the mean, width, and signal yield are freely varied. Combinatorial background is modelled by an exponential function and the slope and yield are determined in the fit to data. The partially reconstructed backgrounds to the left of the signal peak are modelled as in Ref.~\cite{LHCb-PAPER-2023-040}. The results of the fit for the two- and four-body SS decays are shown in Figs.~\ref{fig_global_fit_kpi} and ~\ref{fig_global_fit_kpipipi}, respectively. The yield of the partially reconstructed background in the region above 5230\mevcc is \mbox{$25 \pm 1$} in the \mbox{$D \to K \pi$} decay mode and this is used in the subsequent fit to determine the \CP observables, to determine the residual background yield in other decay channels.

In certain channels, there are further peaking backgrounds under or near the signal peak that need to be accounted for in the fit. 
One is the decay of a $B$ meson to the same final state as the signal, but with no intermediate charm meson. 
This charmless background is reduced by the requirement that the $D$-meson vertex is displaced from that of the $B$ meson. This background is studied by examining events in the region \mbox{1910\textendash1960\mevcc} of $D$-meson reconstructed mass. Significant yields of this background remain in the \mbox{$D\to\pi\pi$} and \mbox{$D\to\pi\pi\pi\pi$} channels.  They are determined by fits to the distributions of reconstructed $B$ mass in the $D$-meson higher mass sidebands, separately for each $B$ charge. The tail parameters are determined from simulated \mbox{$\Bpm \to \pip\pim\Kstarpm$} decays and the mean and width from data. The shape of the background is similar to, but wider than, the signal.

Background from \mbox{$\decay{\Lb}{\Lc(\to p\Km[\pip])\Kstarm}$} decays can contribute under the signal peak in $B$-meson reconstructed mass for the \mbox{$D \to \Kp\Km$} decay mode when the pion from the $\Lc$ is missed in the reconstruction and the proton misidentified as a kaon. The shape of this background is determined from a fit to the reconstructed-mass distribution of simulated \mbox{$\decay{\Lb}{\Lc\Kstarm}$} decays, that are reconstructed in the same way as the signal decays, with an extended Cruijff function used to model it. The yield of the background is determined relative to the signal yield taking into account the relative production ratios of the \Lb and \Bpm hadrons~\cite{LHCb-PAPER-2011-018}, the ratio of branching fractions of the signal and background decays~\cite{PDG2024} and the relative efficiency of selecting the background in comparison to the signal decay mode. This background is assumed to occur equally in each charge category and is approximately 2$\%$ of the signal yield for the $\D\to \Kp\Km$ mode. 

Decays of the type \mbox{$\Bs \to \Dzb \Kbar^{**0}$}, with \mbox{$\Kbar^{**0} \to \Kstarm [\pip]$} and \mbox{$\Kstarm \to \KS\pim$}, where the pion from the $\Kbar^{**0}$ decay is missed and not reconstructed, can appear in the region below the signal peak in $B$-meson reconstructed mass. 
These decays are favoured in the OS modes considered in this measurement, where low signal yields are expected. These backgrounds are suppressed in the SS decay modes and hence the background cannot be determined from that sample. Fast simulation~\cite{Cowan:2016tnm} is used to generate samples of \mbox{$\Bs \to \Dzb \Kbar^{**0}$} decays, where $\Kbar^{**0}$ is one of the following resonances: $\Kbar_1(1270)^0$, $\Kbar_1(1400)^{0}$, $\Kstarb(1410)^{0}$, $\Kstar(1430)^{0}$ and $\Kbar_{2}^{*}(1430)^{0}$. 
The resulting reconstructed mass shape is similar and averaged over the five resonances. To understand the potential yield of this background, the low sideband of reconstructed $B$ mass in the two-body OS decay mode with both charge categories is inspected. 
The expected yield of combinatorial background is subtracted by extrapolating from the region above 5400\mevcc. The expected yield of \mbox{$\B \to \Dstar \Kstar$} decays is estimated from the SS decay. 
The yield is scaled to take into account the efficiency differences of the OS and SS decay modes and the physics expectation of the ratio. This is dependent on the physics parameters of interest; for the purpose of this estimate, the central values of $\gamma$, $r_B$ and $\delta_B$ from Ref.~\cite{LHCb-CONF-2022-003} are used. 
This yield is also subtracted from the data distribution. The remaining distribution is assumed to come solely from the $\Bs$ decay, and is fit to the distribution determined from simulation. The expected number of candidates above 5230\mevcc is approximately 2.5 which is split across both charges. 
Since this yield is very low, the assumptions on the backgrounds used to determine this yield have negligible impact. The background is also scaled for the OS four-body decay mode. The background is not significant in the self-conjugate or SS decay modes and is not considered further. 

\section{Signal yields and observation of  decays  }
\label{sec:yields}

The fit is first performed with the signal yield freely varying. Each decay mode is split by $B$-meson charge, and a simultaneous fit is carried out, where the categories are the $D$-meson decay channel and $B$-meson charge. 
The mass range is reduced to \mbox{5230\textendash5600\mevcc} to remove the partially reconstructed \mbox{$B \to \Dstar \Kstar$} backgrounds. The two main components of the fit are the signal and the combinatorial background. 
The signal is modelled by the same function used in the fit with a wider mass range, where the mean and width are freely varying. The two-body decay categories of the $D$ meson all share the same mean and width. The same is the case for the four-body $D$-decay categories. 
For the three-body decay categories, the signal function is simplified and consists only of the modified Cruijff PDF. The $\mu$ and $\sigma$ parameters are freely varying and shared across the three-body decay categories with the tail parameters determined from simulation. The combinatorial background is modelled by an exponential PDF. The yield of this background is freely varying in each category. The slope of the exponential PDF is shared between the two-, three-, and four-body categories. For the small residual partially reconstructed background, the reconstructed shape is that from the larger mass range fit described in Sec.~\ref{sec:bkg} to the SS decay modes and the yield is determined by scaling by the relative branching fractions of the $D$-meson decays. Backgrounds specific to certain decay channels described in Sec.~\ref{sec:bkg} are also included. 

The results of this fit are given in Table~\ref{table:cpfit_yields}. As expected, there are large yields in the SS two- and four-body decay modes, approximately 2.5 times larger than those in Ref.~\cite{LHCb-PAPER-2017-030}. This is due to the inclusion of data collected in 2017 and 2018. The yield of the $B$ decay where \mbox{$D \to \KS \pi\pi$} is an order of magnitude larger than that analysed at Belle~\cite{belle_prev} and around three times that analysed at BaBar~\cite{babar_prev}. A further fit is performed where the OS decay modes have both charges combined. Using Wilks' theorem~\cite{Wilks:1938dza}, it is found that the statistical significance of an observation for $\pi K$ and $\pi K \pi \pi$ is above five standard deviations. This is the first observation of the suppressed \mbox{$\Bpm \to [\pipm\Kmp]_D \Kstarpm$} and \mbox{$\Bpm \to [\pipm\Kmp\pipm\pimp]_D \Kstarpm$} decays.

\begin{table}[t] \centering
  \caption{Summary of the observed signal yields for each $D$-decay channel and for each $B$-meson charge. The uncertainties are statistical only. The purity in the mass region \mbox{5230\textendash5330\mevcc} is also given for both the charges combined. }
  \label{table:cpfit_yields}
\begin{tabular}{cccc}
\\[-1.8ex]\hline
\hline \\[-1.8ex]
\multicolumn{1}{c}{$D$ decay mode} & \multicolumn{1}{c}{$\Bm$} & \multicolumn{1}{c}{$\Bp$} & Purity (\%) \\ \hline\\[-1.8ex]

$K\pi$ & $2656 \pm 55$  & $2844 \pm 57$ & 94    \\
$KK$ & $\phantom{0}366 \pm 20$ & $\phantom{0}274 \pm 18$  & 90    \\
$\pi\pi$ & $\phantom{0}121 \pm 13$    & $\phantom{00}63 \pm 10$ & 69   \\
$\pi K$   & $\phantom{00}5 \pm 4$  & $\phantom{0}35 \pm 7$ & 58  \\
$K\pi\pi\pi$  & $1665 \pm 44$ & $1783 \pm 45$ & 93  \\
$\pi\pi\pi\pi$ & $\phantom{0}160 \pm 14$  & $\phantom{0}149 \pm 14$  & 77 \\ 
$\pi K\pi\pi$  & $\phantom{0}13 \pm 5$  & $\phantom{0}18 \pm 5$ & 58  \\
$\KS \pi\pi$  & $\phantom{0}279 \pm 18$  & $\phantom{0}268 \pm 18$ & 85  \\
$\KS KK$  & $\phantom{0}29 \pm 6$  & $\phantom{0}40 \pm 7$ & 86  \\ 
\hline
\hline \\[-1.8ex]

\end{tabular}
\end{table}

\section{Determination of $\CP$-violating observables}
\label{sec:cpfit}
The \CP-violating observables are determined through fits to the reconstructed-mass distributions where the events are separated by category. The categories are each $D$-decay mode, separated into $\Bp$ and $\Bm$ candidates. The three-body decay modes are further separated into the defined phase-space regions on the Dalitz plot.

 The relative signal yields in this simultaneous fit to the different mass distributions are expressed in terms of the $\CP$-violating parameters, which allows the direct determination of their statistical correlations and eases the subsequent determination of common systematic uncertainties. 
 Corrections must be included in the fit to take into account $\Bpm$ production asymmetries and detection asymmetries for $\Kpm$ and $\pipm$ due to differences in the interaction with the material as they pass through the \lhcb detector. 
 The $\Bpm$ production asymmetry, \mbox{$A_{\Bp} = (+0.028 \pm 0.068) \%$}, is taken from Ref.~\cite{LHCb-PAPER-2020-036}, which is in agreement with that in Ref.~\cite{LHCb-PAPER-2016-054}. 
 The kaon detection asymmetry is mainly due to the nuclear interaction length difference between $\Kp$ and $\Km$. This is corrected using \mbox{$A_{\Kp\pim} = (-0.960 \pm 0.134)$\%}, which is defined such that a higher $\Kp \pim$ detection efficiency leads to a positive asymmetry, and pion detection asymmetry \mbox{$A_{\pim} = (-0.064 \pm 0.019)$\%}, where the uncertainty includes systematic effects~\cite{LHCb-PAPER-2016-054}. 

In the case of the two- and four-body decay modes the signal yield in each decay mode and each charge is reparametrised in terms of the \CP-violating observables, the production and detection asymmetries, the yield of the SS decay that has the same $D$-decay final-state multiplicity and efficiency ratios. These efficiency ratios take into account the differences in selection between the signal channel and the relevant SS decay channel. The efficiency is determined through simulation except for the efficiency of PID requirements which is determined using data. 

In the case of the three-body decay modes, there are two separate free normalisation parameters for the two $B$-meson charges. This means that the result is independent of production and detection asymmetries. The signal yields in each bin are parametrised using Eq.~\ref{Eq:B-}, where the two normalisation parameters are set up to be the total $\Bp$ and $\Bm$ yield integrated over the phase space for each decay channel. The $F_i$ parameters are those determined in Ref.~\cite{LHCb-PAPER-2020-019}, including their reparametrisation to ensure fit stability. The values of $c_i$ and $s_i$ are determined from a combination of \besiii~\cite{BESIII, bes3kk} and CLEO~\cite{CLEO} measurements and are fixed in the fit. The PDF in each category is the same as for the data combined across phase space. The tail parameters are fixed from simulation. The mean mass and width are shared across the categories. The slope of the combinatorial background is fixed from the fit integrated over phase space and is the same in each phase-space region. However, the yield is determined in the fit for each category. The yield of partially reconstructed background is determined from scaling the yield observed in the \mbox{$D\to K\pi$} SS decay. In a $\Bm$ decay it is assumed that the partially reconstructed background contains a $\Dz$, and hence the yield of this background is distributed over the phase space with this assumption. As there are approximately six candidates for partially reconstructed background in \mbox{$D\to\KS\pi\pi$} and one candidate in \mbox{$D\to \KS KK$}, the impact of this assumption is negligible. Hence, the free parameters that describe the data are the mean and width of the signal, the total yield of $\Bp$ and $\Bm$ for each decay channel, the yield of combinatorial background in each category, and the \CP-violating observables $x_\pm$, $y_\pm$.

The fits for all decay modes are tested through extensive pseudoexperiments to assess any fit bias. No bias is found and the uncertainties are found to be well estimated for the two- and four-body decay modes. In the case of the three-body decay modes there is evidence of a small bias up to 7\% of the statistical uncertainty. The source of the bias is determined to be due to the very low yields that occur in some categories. The fit results are corrected for this small bias.

\begin{figure}
    \centering
    \includegraphics[width=0.85\linewidth]{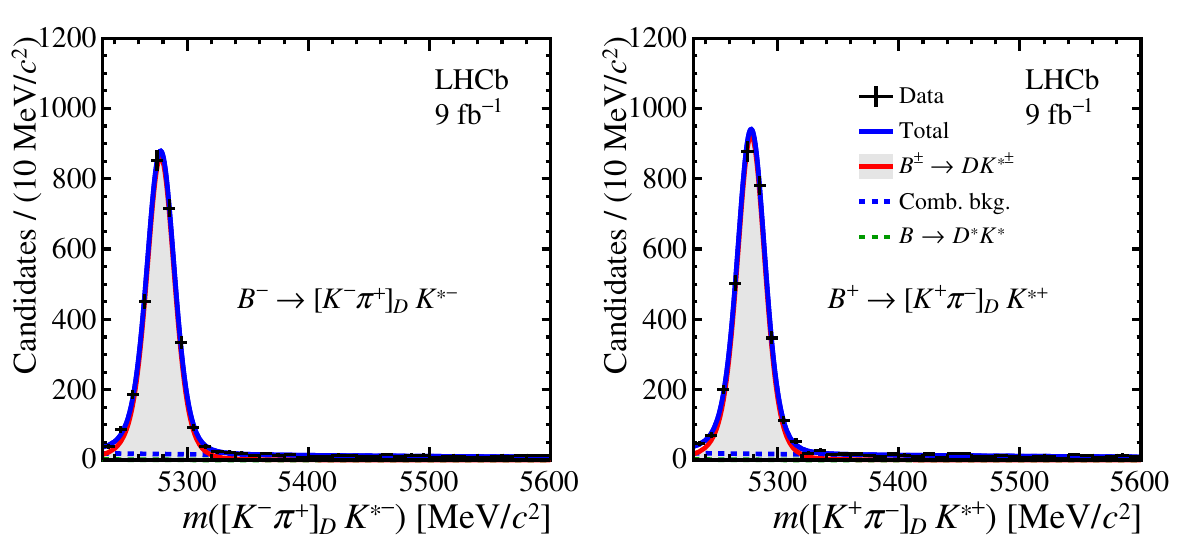}
    \includegraphics[width=0.85\linewidth]{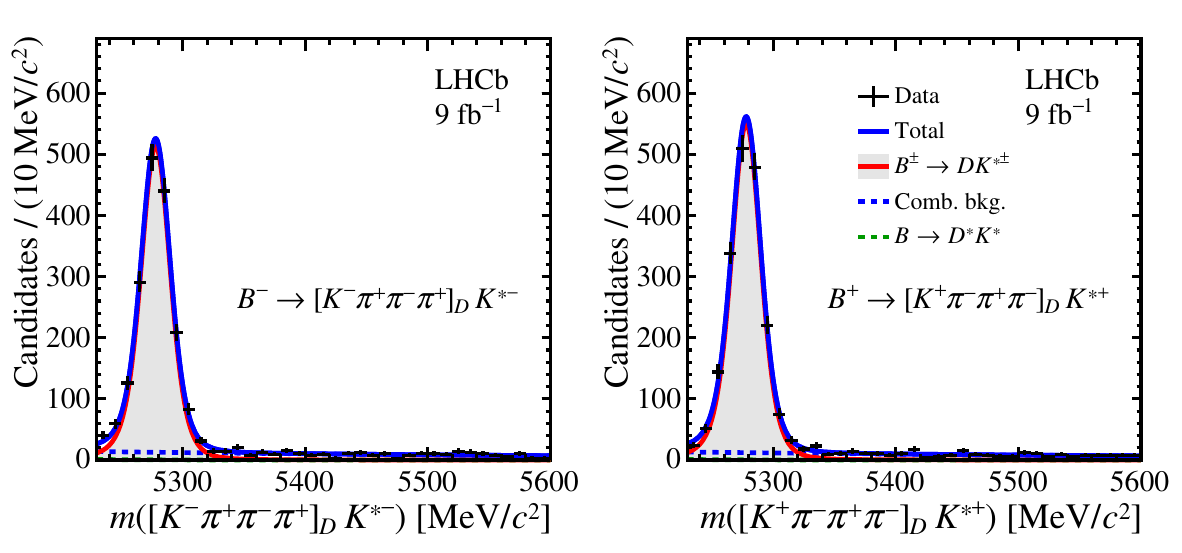}
    \caption{Reconstructed-mass distributions for the favoured (top) $K\pi$ and (bottom) $K \pi \pi \pi$ modes for (left) $B^-$ and (right) $B^+$ decays, obtained with data corresponding to the full Run 1 and Run 2 datasets. Fit result is also shown.}
    \label{fig_asym_res1}
\end{figure}

\begin{figure}
    \centering
    \includegraphics[width=0.91\linewidth]{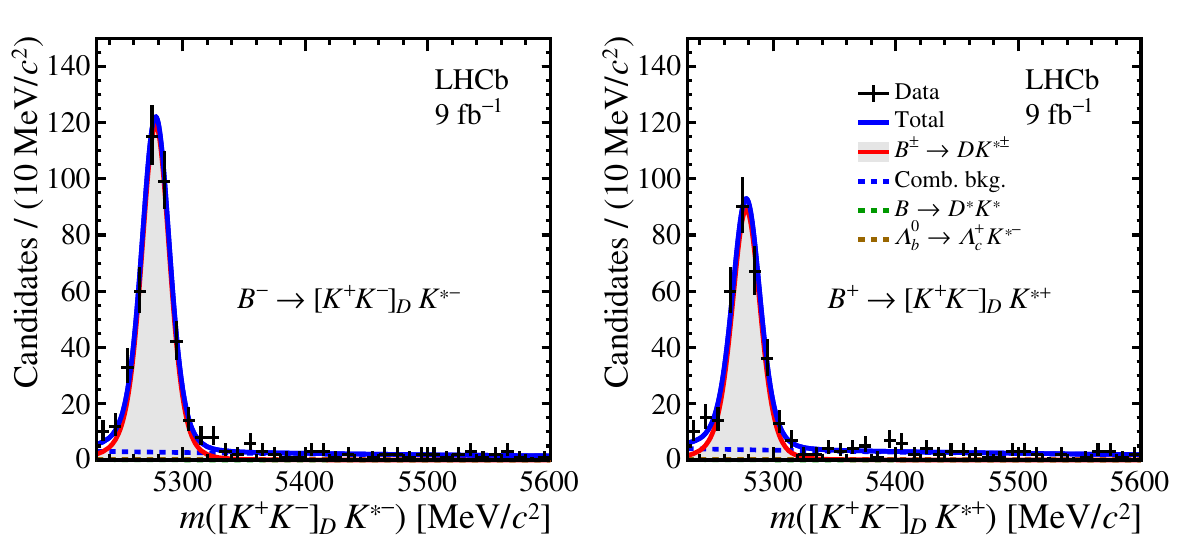}
    \includegraphics[width=0.91\linewidth]{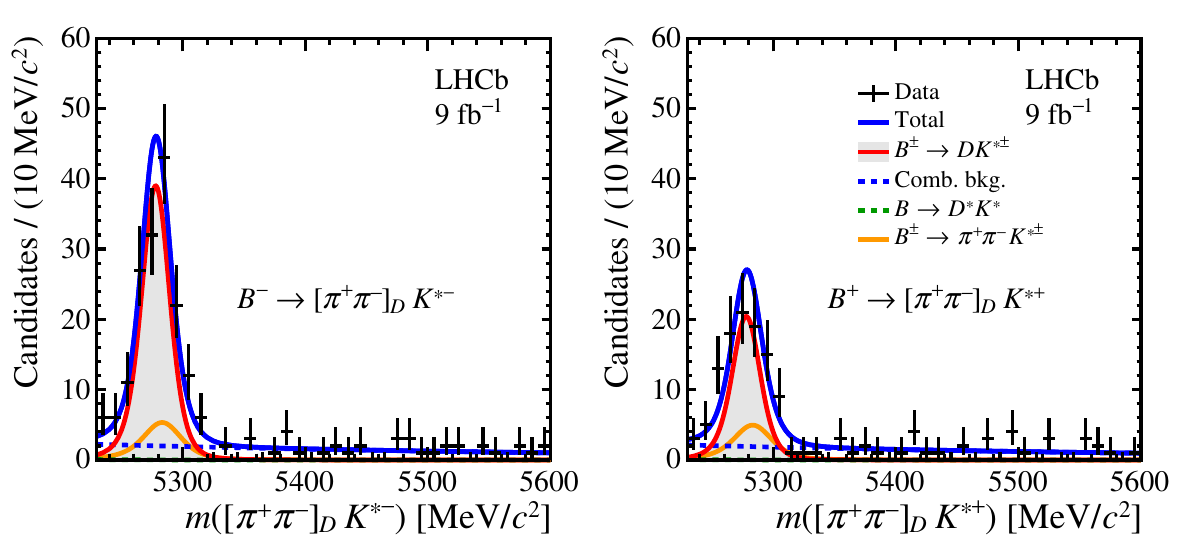}
    \includegraphics[width=0.91\linewidth]{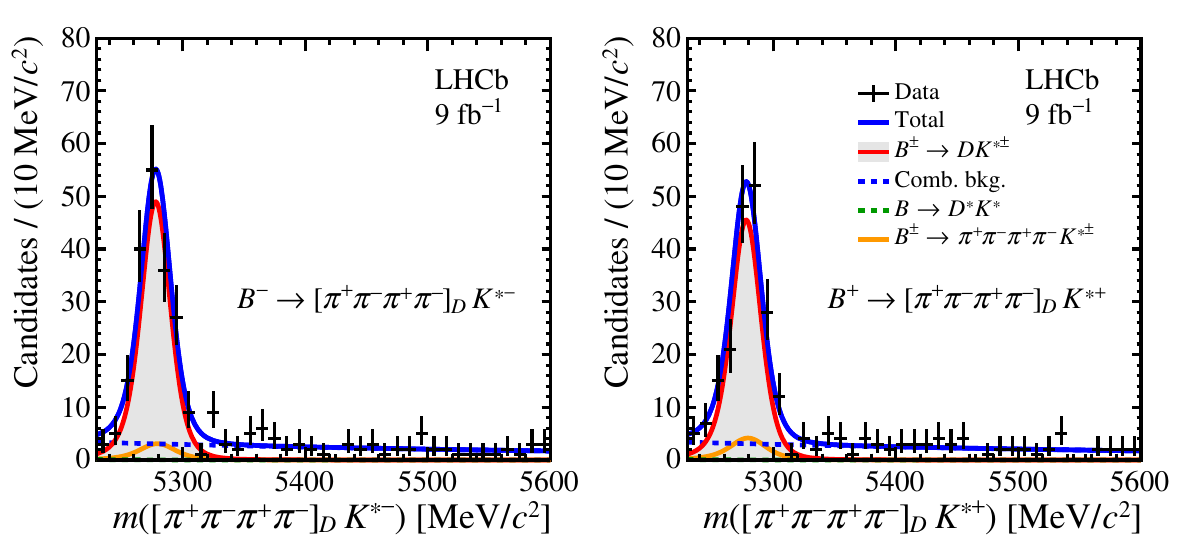}
    \caption{Reconstructed-mass distributions for the (top row) $KK$, (middle row) $\pi \pi$ and (bottom row) $\pi \pi \pi \pi$ modes for (left) $B^-$ and (right) $B^+$ decays, obtained with data corresponding to the full Run 1 and Run 2 datasets. Fit result is also shown.}
    \label{fig_asym_res2}
\end{figure}

\begin{figure}
    \centering
    \includegraphics[width=0.91\linewidth]{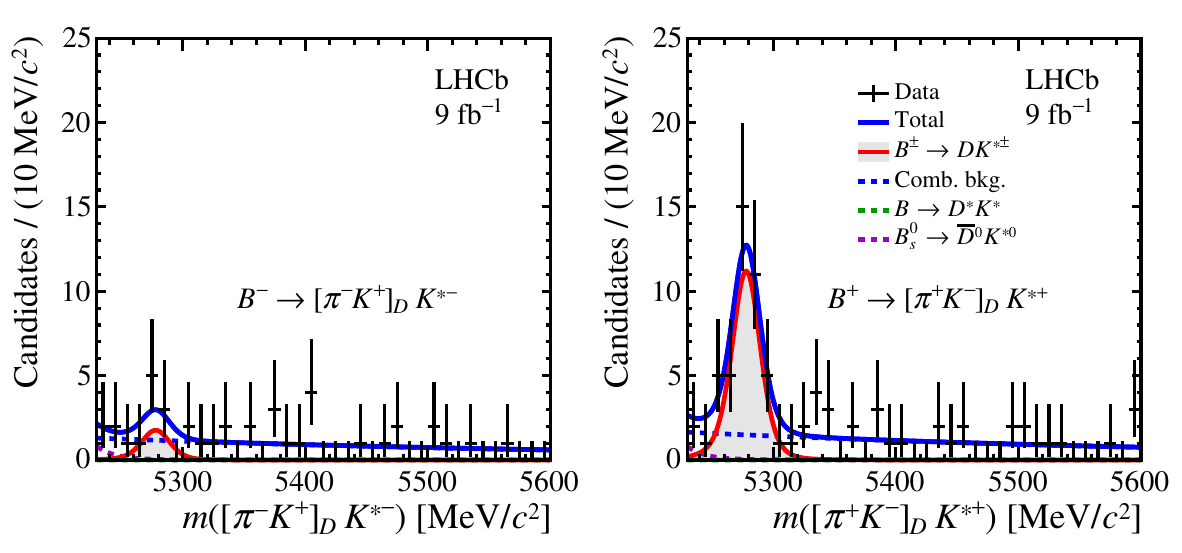}
    \includegraphics[width=0.91\linewidth]{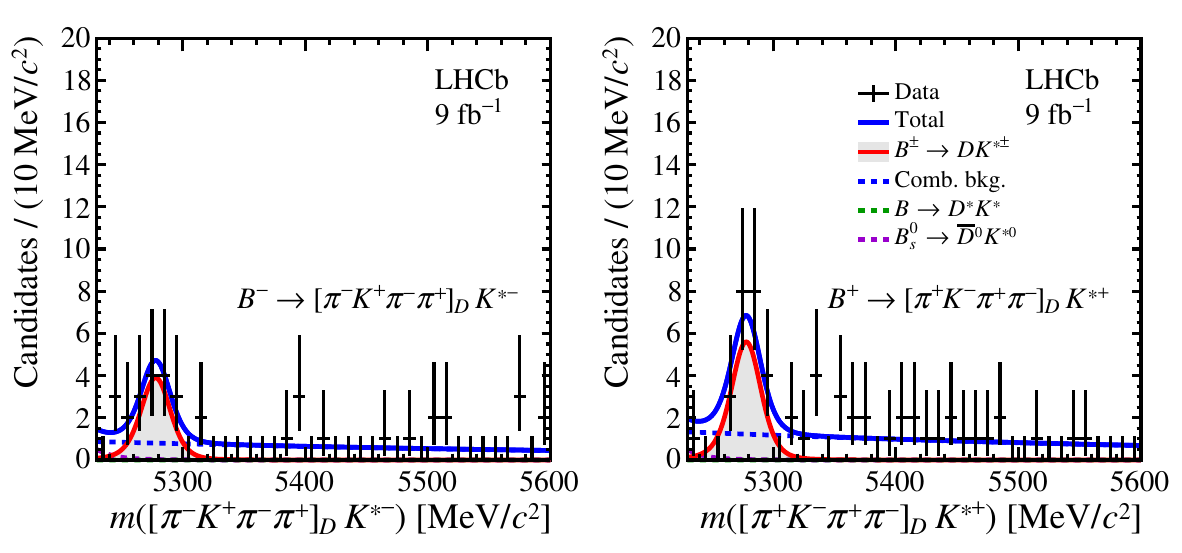}
    \caption{Reconstructed-mass distributions for the suppressed (top row) $\pi K$ and (bottom row) $\pi K \pi \pi$ modes for (left) $B^-$ and (right) $B^+$ decays, obtained with data corresponding to the full Run 1 and Run 2 datasets. Fit result is also shown.}
    \label{fig_asym_res3}
\end{figure}

\begin{figure}
    \centering
    \includegraphics[width=0.65\linewidth]{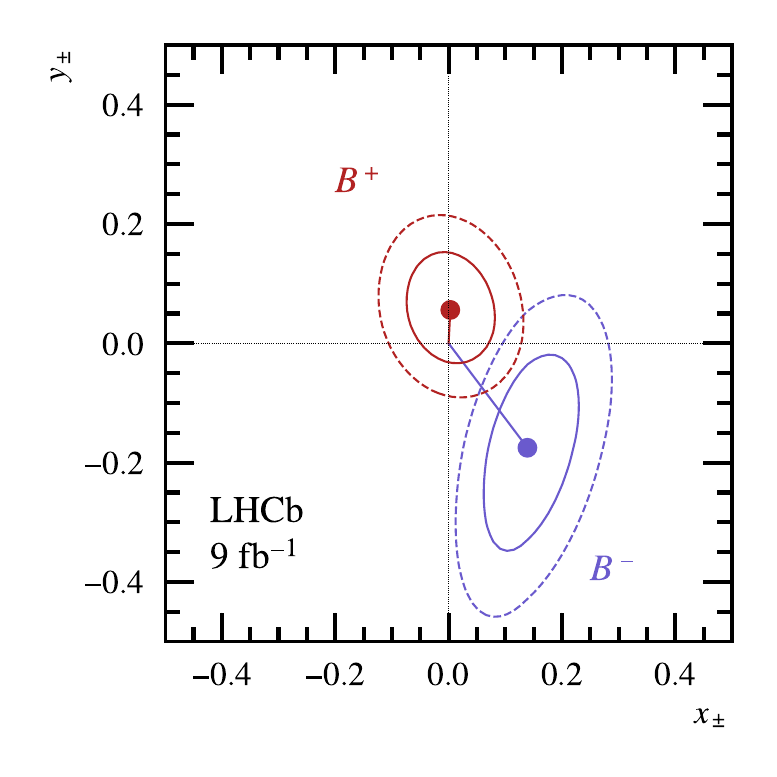}
    \caption{Central values for the measured
($x_{\pm}, y_{\pm}$) parameters along with two-dimensional confidence regions of 68\% and 95\%, considering statistical uncertainty. The red (blue)
contours correspond to the observables related to $B^+$ ($B^-$) decays with three-body $D$ modes.}
    \label{fig_likelihood}
\end{figure}

\begin{figure}
    \centering
    \includegraphics[width=0.75\linewidth]{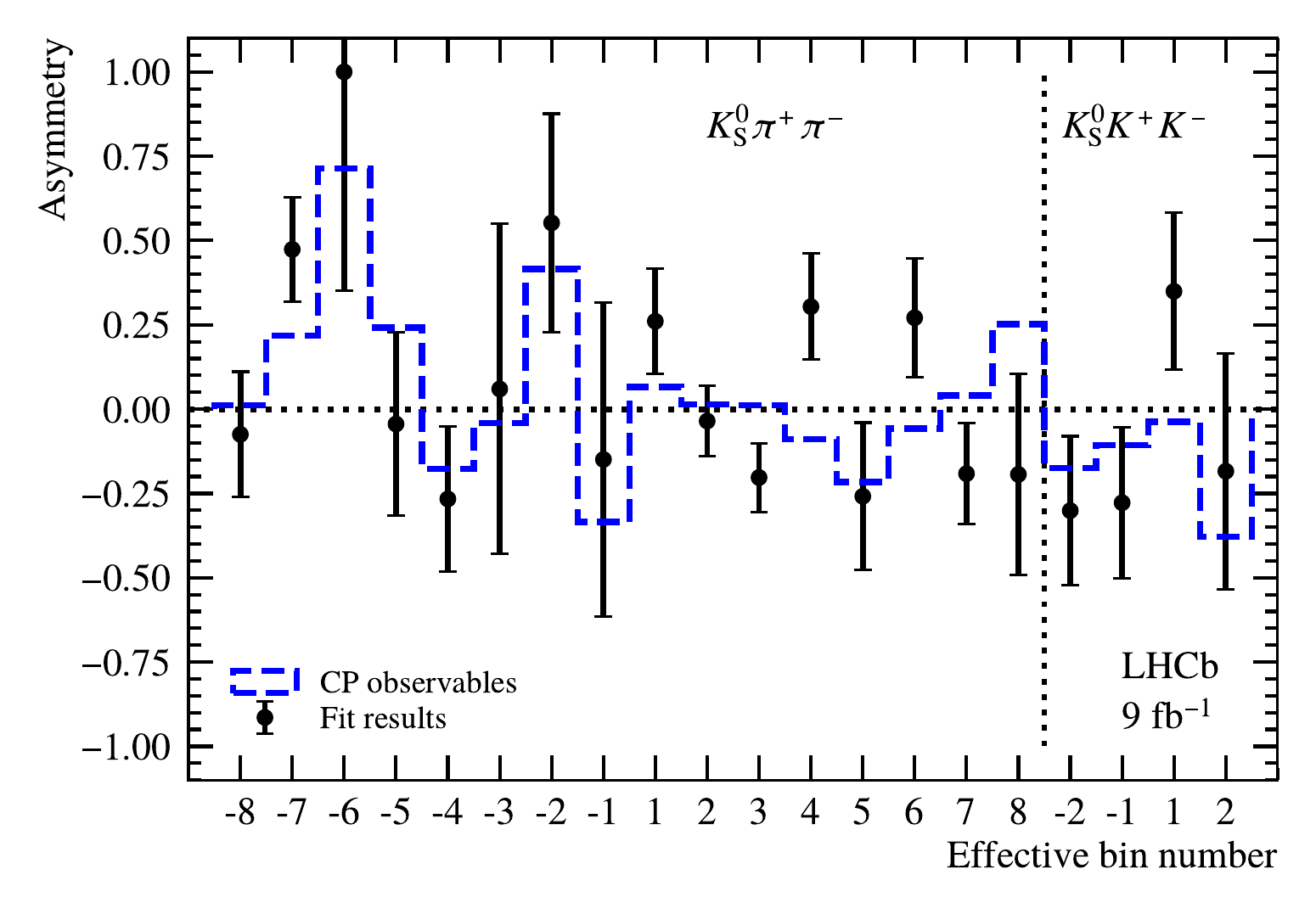}
    \caption{Raw asymmetry in each effective bin pair. It is determined using the fitted ${\ensuremath{C\!P}}$-violating observables (blue histogram)
and the results of an alternative fit where the signal yield in each Dalitz plot bin is a free
parameter (black points, with statistical uncertainties).}
    \label{fig_asym_bins}
\end{figure}

The results of the fit for the two- and four-body decay modes are shown in Figs.\mbox{~\ref{fig_asym_res1}$\textendash$\ref{fig_asym_res3}}, with numerical results provided in Sec.~\ref{sec:results}. There are small asymmetries within the SS sample, as expected, with noticeably larger values for the OS and (quasi)-\CP-eigenstate decay channels, although not all are statistically significant. The results of the three-body decay modes are visually presented in Fig.~\ref{fig_likelihood}, where the $x_{\pm}$ and $y_\pm$ are shown. The opening angle between the two vectors joining the central values to the origin is $2\gamma$. To investigate the goodness-of-fit in the three-body decay channels, a further fit is run, where the signal yield is not parametrised in terms of the fit observables described above, but freely varying in each category. Asymmetries are calculated and shown in Fig.~\ref{fig_asym_bins}. The asymmetries are calculated in effective bin pairs, labelled $i$, that compare the yield of \Bm decays in a bin $i$ with the yield of \Bp decays in a bin $-i$. Also shown are the expected asymmetries given the fitted values of $x_\pm$ and $y_\pm$. The good agreement between the two approaches demonstrates that the system of equations used to fit the data is reasonable. 

\section{Systematic uncertainties}
\label{sec:syst}
Different sources of systematic uncertainties contribute to the measurement. Two methods are used to compute the systematic uncertainties on the \CP observables. To evaluate the uncertainty due to parameters that are kept constant in the \CP fit, the data fit is repeated thousands of times while varying the given parameter according to a Gaussian function whose width corresponds to the uncertainty on the parameter. Where more than one parameter is varied (\eg the $c_i$), the correlations between them are taken into account. The systematic uncertainty for each \CP observable is taken to be the standard deviation of the fit parameter distribution from these many fits. Other systematic uncertainties are evaluated by generating thousands of pseudoexperiments with an alternative model and fitting back with the default fit. In this case, the systematic uncertainty is computed as the difference between the generated value and the mean of the fit parameter distribution.

In the $\CP$ fit involving two- and four-body decays, uncertainties are assigned for the fixed values of production and detection asymmetries, branching fractions, and the ratios of selection efficiencies. Systematic uncertainties related to the signal mass model are assessed by considering the impact of the fixed parameters in the signal shape.

To assess the assumption that the slope of the combinatorial background is the same in decay modes with the same number of final-state particles, the high sideband region of $B$-meson reconstructed mass is fitted independently for each $D$-meson decay mode. Pseudoexperiments are generated with the result of this fit, but fitted back with the nominal model to determine the corresponding uncertainty.
Uncertainties in the yields of the partially reconstructed backgrounds that remain present above 5230\mevcc of the reconstructed mass are estimated from varying the yields and shape in the initial fit. The assumption of no \CP violation in these decays does not contribute to any significant systematic effects. The systematic uncertainties for the other backgrounds are also determined by variation of the yield and shape of each background. Due to the nature of the simultaneous fit, those backgrounds present only in certain modes have small effects on \CP parameters that are primarily determined from other decay modes. 

The summary of systematic uncertainties is given in Table~\ref{table:summary_syst}. In all modes except \mbox{$D \to \pi\pi\pi\pi$}, the systematic uncertainty is less than 50\% of the statistical one. For the \mbox{$D\to\pi\pi$} and \mbox{$D\to \pi\pi\pi\pi$} decay modes, the impact of the uncertainties in the charmless decays is significant. Uncertainties on the branching fractions have a large impact in the ratio observables. 

In the three-body decay modes, the leading systematic uncertainty on $y_\pm$ arises from the input parameters $c_i$ and $s_i$ . The $F_i$ values are also fixed in the fit and determined from \mbox{$\Bpm \to D \pipm$} decays. Small differences in the relative efficiency over the Dalitz plot of this control channel and the signal channel are determined in simulation. The changes are due to a looser requirement on the momentum of \textit{long} $\KS$ candidates in \mbox{$\btodkstar$} decays at the initial stage of the selection. These changes are used to modulate the $F_i$ values with which psuedoexperiments are generated. They are fitted with the nominal $F_i$ values to determine the associated systematic uncertainty. This is the leading source of uncertainty on $x_\pm$. 

\begin{landscape}
\vspace*{\fill}
\begin{table}[!htbp] \centering
\footnotesize
  \caption{Summary of systematic uncertainties for two- and four-body modes ($\times 10^{-2}$). Contributions which are less than 0.1\% of the statistical uncertainty are not shown.}
  \label{table:summary_syst}
\begin{tabular}{lcccccccccccc}
\\[-1.8ex]\hline
\hline \\[-1.8ex]
\multicolumn{1}{l}{ } & \multicolumn{1}{c}{$A_{SS}^{K\pi}$} & \multicolumn{1}{c}{$A_{\CP}^{KK}$} & \multicolumn{1}{c}{$A_{\CP}^{\pi \pi}$} & \multicolumn{1}{c}{$A_{OS}^{\pi K}$} & \multicolumn{1}{c}{$R_{\CP}^{KK}$} & \multicolumn{1}{c}{$R_{\CP}^{\pi \pi}$} & \multicolumn{1}{c}{$R_{OS}^{\pi K}$} & \multicolumn{1}{c}{$A_{SS}^{K\pi\pi\pi}$} & \multicolumn{1}{c}{$A_{\CP}^{\pi \pi \pi \pi}$} & \multicolumn{1}{c}{$A_{OS}^{\pi K \pi \pi}$} & \multicolumn{1}{c}{$R_{\CP}^{\pi \pi \pi \pi}$} & \multicolumn{1}{c}{$R_{OS}^{\pi K \pi \pi}$} \\ \hline\\[-1.8ex] 
Asymmetry corrections & 0.17\phantom{0} & 0.072 & 0.067 & \phantom{0}0.078 & -- & -- & -- & 0.17\phantom{0} & 0.073 & \phantom{0}0.16\phantom{0} & -- & -- \\
Branching fractions & -- & -- & -- & -- & 0.88\phantom{0} & 1.2\phantom{00} & -- & -- & -- & -- & 3.5\phantom{0} & -- \\
Selection efficiencies & -- & -- & -- & -- & 0.87\phantom{0} & 0.76\phantom{0} & 0.0024 & -- & -- & -- & 1.2\phantom{0} & 0.0047 \\
PID efficiencies & -- & -- & -- & -- & 0.22\phantom{0} & 0.23\phantom{0} & -- & -- & -- & -- & 0.36 & -- \\
Signal shape & -- & -- & 0.046 & \phantom{0}0.067 & 0.20\phantom{0} & 0.26\phantom{0} & 0.0011 & -- & 0.020 & \phantom{0}0.069 & 0.31 & 0.0021 \\
Combinatorial shape & 0.034 & 0.053 & 0.14\phantom{0} & \phantom{0}2.6\phantom{00} & 0.30\phantom{0} & 0.29\phantom{0} & 0.021\phantom{0} & 0.014 & 0.22\phantom{0} & \phantom{0}0.097 & 0.14 & 0.0071 \\
Part. reco. background & -- & -- & -- & \phantom{0}0.16\phantom{0} & 0.072 & 0.12\phantom{0} & 0.0043 & -- & -- & -- & -- & -- \\
Charmless background & -- & -- & 4.9\phantom{00} & \phantom{0}0.034 & -- & 4.5\phantom{00} & -- & -- & 2.9\phantom{00} & -- & 3.0\phantom{0} & -- \\
\Lb background & -- & -- & 0.016 & \phantom{0}0.044 & 0.030 & 0.039 & -- & -- & -- & -- & -- & -- \\
\Bs background & 0.046 & 0.011 & 0.38\phantom{0} & \phantom{0}1.1\phantom{00} & 0.020 & 0.032 & 0.0093 & 0.038 & 0.12\phantom{0} & \phantom{0}0.54\phantom{0} & 0.27 & 0.0054 \\ \hline
Total systematic & 0.18\phantom{0} & 0.09\phantom{0} & 4.9\phantom{00} & \phantom{0}2.8\phantom{00} & 1.3\phantom{00} & 4.7\phantom{00} & 0.02\phantom{00} & 0.17\phantom{0} & 2.9\phantom{00} & \phantom{0}0.5\phantom{00} & 4.8\phantom{0} & 0.01\phantom{00} \\ \hline
Statistical & 1.4\phantom{00} & 4.0\phantom{00} & 9.0\phantom{00} & 16.4\phantom{00} & 5.0\phantom{00} & 9.0\phantom{00} & 0.19\phantom{00} & 1.8\phantom{00} & 6.0\phantom{00} & 21.8\phantom{00} & 7.0\phantom{0} & 0.26\phantom{00} \\ \hline
\hline \\[-1.8ex]
\end{tabular}
\end{table}
\vspace*{\fill}
\end{landscape}

A systematic uncertainty for the fixed value of $\kappa$ is determined by generating pseudodatasets with $\kappa \pm \sigma_{\kappa}$, where $\sigma_{\kappa}$ is the uncertainty associated with it as determined in Ref.~\cite{LHCb-PAPER-2017-030}, and fitting them with the nominal value of 0.95. The largest deviation in the \CP observables is taken as the systematic uncertainty. 
The nonuniform efficiency over the Dalitz plot can lead to a small difference between the measured values of $c_i$ and $s_i$, and efficiency-corrected values. The size of the correction is estimated using simulation to provide the efficiency profile and the decay models from Ref.~\cite{BelleBabarModel} for \mbox{$D\to \KS\pi\pi$} and Ref.~\cite{Babar_KsKK} for \mbox{$D \to \KS KK$}, and fits to pseudodatasets are used to determine the systematic uncertainty. 

Migration between the Dalitz plot bins can occur due to resolution effects on the momentum of final-state particles. To first order, these effects are incorporated into the $F_i$ values. However, second-order effects arise due to the differences in \CP violation in \mbox{$\Bpm \to D \pipm$} and \mbox{$\Bpm \to D \Kstarpm$} decays. These are determined by using the momentum resolution in simulation, the \CP violation observables of \mbox{$\btodpi$}~\cite{LHCb-PAPER-2021-033}, \CP violation observables of \mbox{$\btodkstar$} determined in this analysis, and the \D-decay model from Ref.~\cite{BelleBabarModel}. The observed differences between the two modes are used to generate pseudoexperiments which are then fit with the nominal procedure to assign the systematic uncertainty due to momentum resolution.  
The impact of the signal model is investigated by changing the fit function to a modified Crystal Ball~\cite{Skwarnicki:1986xj} with different widths on both sides. This has an almost negligible impact. Removing the partially reconstructed background entirely has no impact on the fit results and is not considered further. Corrections to the observed bias in the fit model to measure the \CP observables are obtained through pseudoexperiments. The dependence of this bias correction on the values of $\gamma$, $\delta_B$, and $r_B$ used in these simulation studies is investigated by generating alternate pseudoexperiments with varied values of $\gamma$, $\delta_B$, and $r_B$. The change in bias correction values is taken as the systematic uncertainty. A summary of the systematic uncertainties in the three-body decay modes is given in Table~\ref{tab:syst_sum}. The total systematic uncertainties are approximately 35$\%$ of the statistical uncertainties.

\begin{table}[tb!]
\centering
\caption{Summary of systematic uncertainties for three-body modes ($\times 10^{-2}$).}
\label{tab:syst_sum}
\begin{tabular}{lcccc}
\\[-1.8ex]\hline
\hline \\[-2.0ex]
 & $\sigma(x_+)$ & $\sigma(y_+)$ & $\sigma(x_-)$ & $\sigma(y_-)$  \\

\hline
$c_i, s_i$ uncertainty & 0.4 & 1.9 & 0.9 & \phantom{0}3.9 \\
\hline
$F_i$ inputs & 1.5 & 0.4 & 1.7 & \phantom{0}0.4\\
Value of $\kappa$ & 0.8 & 0.4 & 0.6 & \phantom{0}0.8\\
Efficiency correction to $c_i, s_i$ & 0.0 & 0.0 & 0.2 & \phantom{0}0.6\\
Bin migration & 0.4 & 0.2 & 0.3 & \phantom{0}0.4 \\
Mass model & 0.1 & 0.1 & 0.1 & \phantom{0}0.3\\
Bias correction & 0.4 & 0.6 & 0.3 & \phantom{0}0.6 \\
\hline
Total systematic & 1.8 & 2.1 & 2.1 & \phantom{0}4.1\\
\hline
Statistical & 5.2 & 6.4 & 6.0 & 11.4 \\
\hline
\hline

\end{tabular}
\end{table}

\section{Results and interpretation}
\label{sec:results}

The final results of \CP observables for all the decay modes with statistical and systematic uncertainties are:

\begin{alignat*}{3}
A_{SS}^{K\pi} &= -0.024 &&\pm 0.014 &&\pm 0.002, \\
A_{\CP}^{KK} &= \phantom{-}0.14 &&\pm 0.04 &&\pm 0.001, \\
A_{\CP}^{\pi \pi} &= \phantom{-}0.31 &&\pm 0.09 &&\pm 0.05 ,\\
A_{OS}^{\pi K} &= -0.73 &&\pm 0.16 &&\pm 0.03, \\
R_{\CP}^{KK} &= \phantom{-}1.10 &&\pm 0.05 &&\pm 0.01, \\
R_{\CP}^{\pi \pi} &= \phantom{-}0.96 &&\pm 0.09 &&\pm 0.05, \\
R_{OS}^{\pi K} &= \phantom{-}0.0098 &&\pm 0.0019 &&\pm 0.0002, \\
A_{SS}^{K\pi\pi\pi} &= -0.024 &&\pm 0.018 &&\pm 0.002, \\
A_{\CP}^{\pi \pi \pi \pi} &= \phantom{-}0.04 &&\pm 0.06 &&\pm 0.03, \\
A_{OS}^{\pi K \pi \pi} &= -0.19 &&\pm 0.22 &&\pm 0.01, \\
R_{\CP}^{\pi \pi \pi \pi} &= \phantom{-}1.05 &&\pm 0.07 &&\pm 0.05, \\
R_{OS}^{\pi K \pi \pi} &= \phantom{-}0.0118 &&\pm 0.0026 &&\pm 0.0001,
\end{alignat*}

\begin{alignat*}{4}
x_+ &= \phantom{-}0.003&&^{+0.051}_{-0.052} &&\pm 0.018 &&\pm 0.004, \\
y_+ &= \phantom{-}0.056&&^{+0.059}_{-0.064} &&\pm 0.009 &&\pm 0.019, \\
x_- &= \phantom{-}0.139&&^{+0.051}_{-0.060} &&\pm 0.019 &&\pm 0.009, \\
y_- &= -0.175&&^{+0.114}_{-0.103} &&\pm 0.013 &&\pm 0.039.
\end{alignat*}

In the case of the observables $x_\pm$ and $y_\pm$ the third uncertainty represents the propagated uncertainty from the external inputs of $c_i$ and $s_i$. The correlation matrices of the statistical and systematic uncertainties are provided in the Appendix. The correlations related to the strong-phase inputs are given separately.


The \CP-violating observables from the two- and four-body decays are consistent with the previous measurement made on a partial dataset~\cite{LHCb-PAPER-2017-030}. In comparison to the \mbox{$\Bpm \to D \Kpm $} decay~\cite{LHCb-PAPER-2020-019}, the statistical precision on the \CP-violating observables is poorer. This is due to the lower reconstruction efficiency of a $\Kstarpm$ meson compared to a $\Kpm$ meson. However, the \mbox{$\btodkstar$} decays provide good sensitivity due to a larger purity. In particular, there is no counterpart of misidentified background from \mbox{$\Bpm \to D\pipm$} as found in reconstructed \mbox{$\Bpm \to D \Kpm$} candidates. In addition, the separation in reconstructed mass between the signal channel and the partially reconstructed background is larger in this channel than in \mbox{$\Bpm \to D \Kpm$}, which further improves the purity. 

\begin{figure}[th!]
    \centering
    \includegraphics[width=0.49\linewidth]{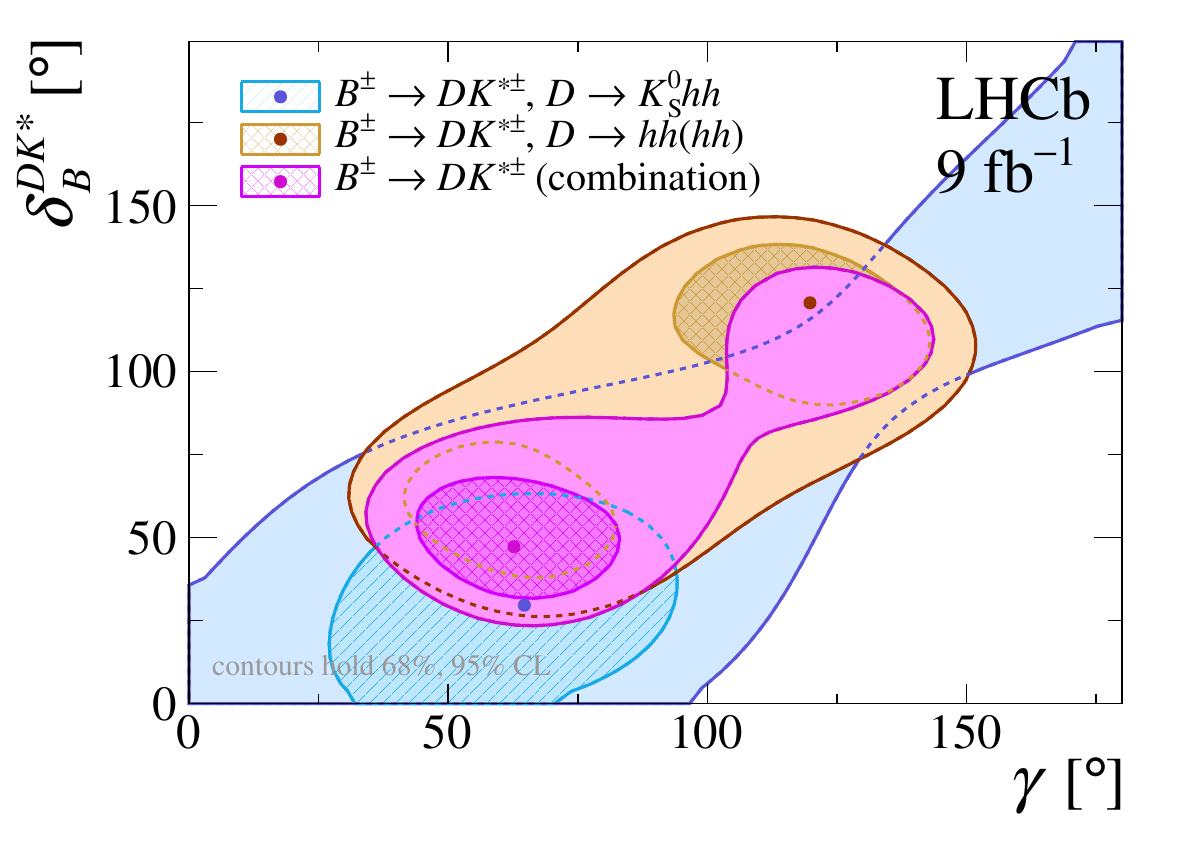}
    \includegraphics[width=0.49\linewidth]{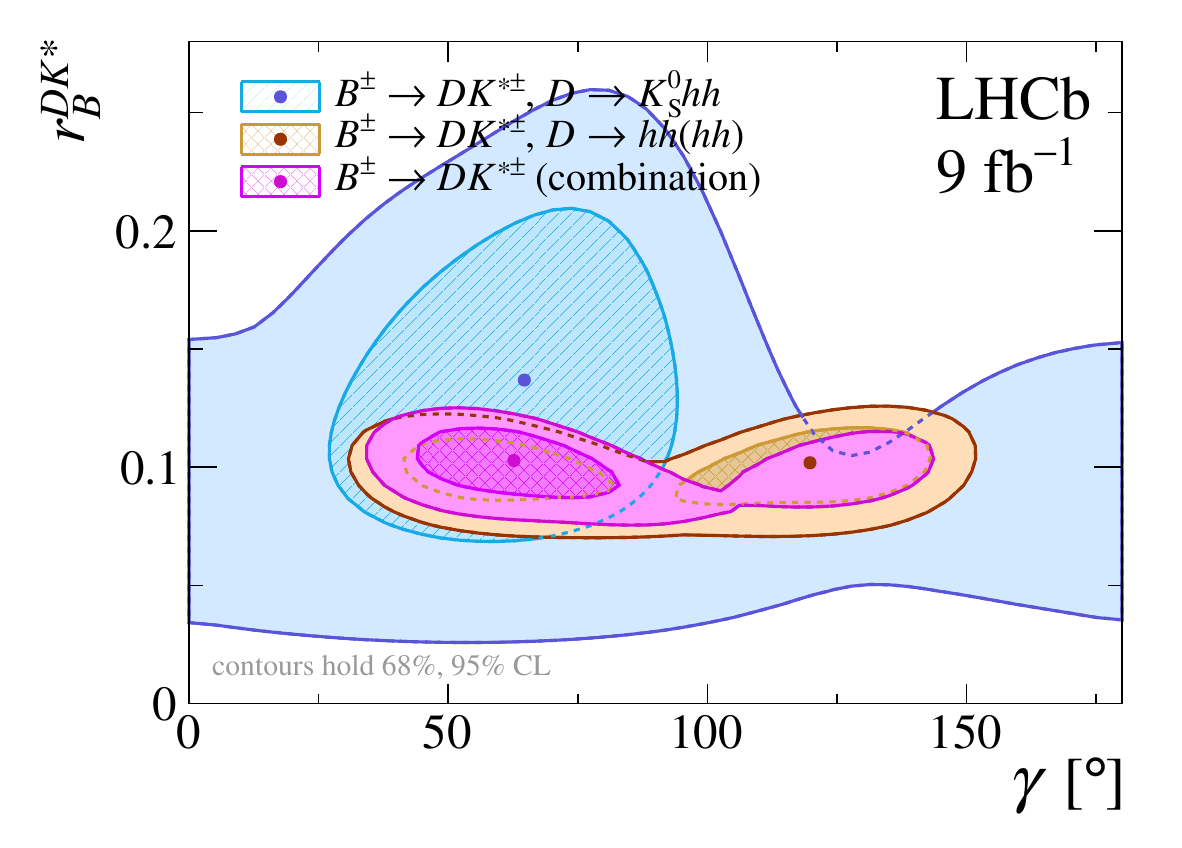}
    \caption{Two-dimensional physics parameter scans showing the contours from the combination with profile-likelihood method. Yellow, blue and magenta contours show the results for two- and four-body, three-body and all $D$-decay modes, respectively. The hatched and solid regions correspond to 68\% and 95\% confidence level, respectively.}
    \label{fig_combo_contours_all}
\end{figure}

The measured \CP-violating observables are related to the physics parameters of interest via Eqs.\mbox{~\ref{eq:R_kk}\textendash\ref{eq:A_adsfav}} and \ref{Eq:B-}  in Sec.~\ref{sec:method}. The interpretation is performed using the {\tt GammaCombo} package \cite{GammaCombo,LHCb-PAPER-2021-033}, which can implement a simple profile-likelihood method or the Feldman--Cousins approach~\cite{FC} combined with a ``plugin'' method~\cite{Plugin}. The correlations of statistical and systematic uncertainties for the measured \CP-violating observables are accounted for. The $D$-decay parameters $r_D^{K\pi}$ and $\delta_D^{K\pi}$ are inputs from Ref.~\cite{LHCb-CONF-2022-003}, while $r_D^{K 3\pi}$, $\delta_D^{K 3\pi}$, and $\kappa^{K 3\pi}$ are taken from a combination of results from  \besiii, \cleo-c and \lhcb collaborations~\cite{BESIII_Dparams}. Another required input is the $\CP$-even fraction $F_{\CP}$ for the decay \mbox{$D\to\pi\pi\pi\pi$} which is measured at \besiii~\cite{BESIIIFPlus4Pi}. The combination of the \CP-violating observables from the two- and four-body decay modes leads to two solutions for $\g$ in the range \mbox{0\textendash180} degrees as shown in Fig.~\ref{fig_combo_contours_all}. The three-body decay modes lead to a single solution, although in two-dimensional parameter space no values of $\gamma$ are excluded at the 95\% confidence level. In the combination shown in Fig.~\ref{fig_combo_1d_gamma_all}, there is a single solution which is consistent with the current world average of $\gamma$. The following values of the physics parameters are found using the profile-likelihood method:  

\begin{figure}[t!]
    \centering
    \includegraphics[width=0.55\linewidth]{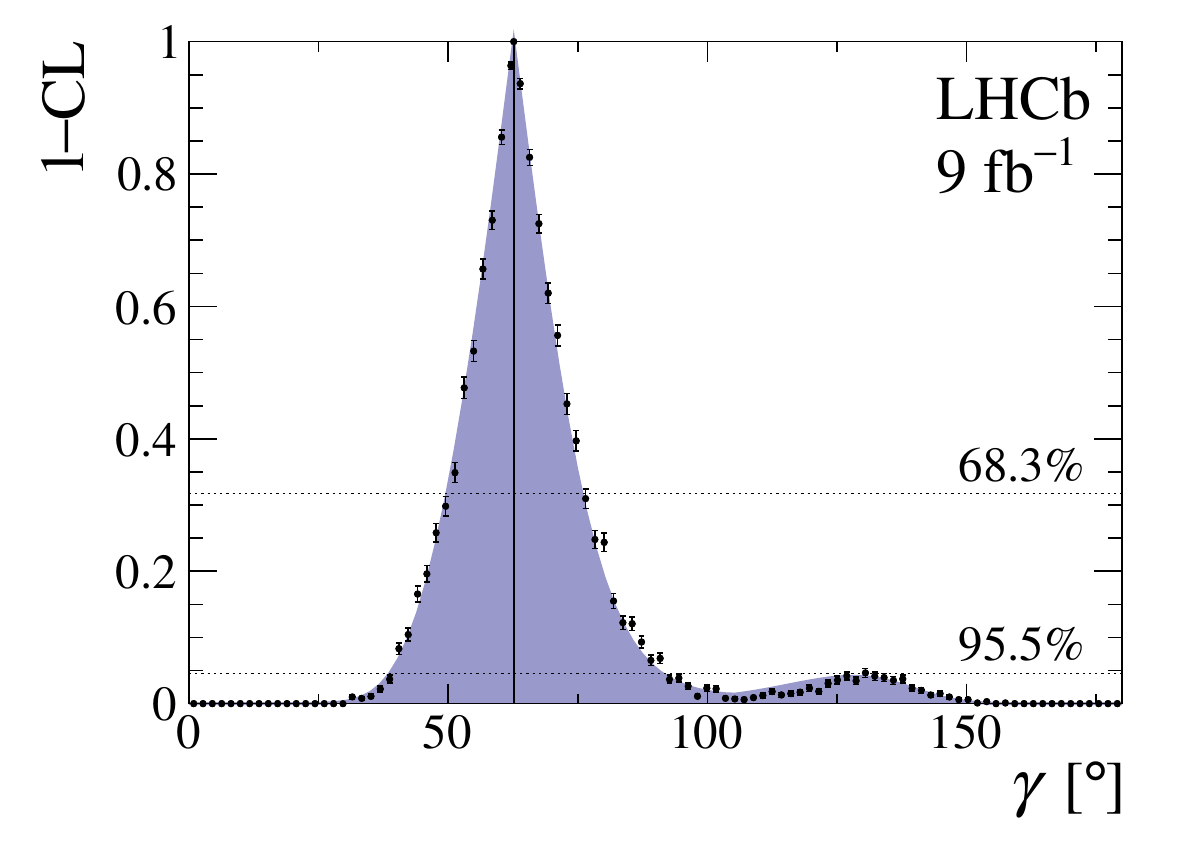}
    \caption{One-dimensional scan on the confidence limits (CL) for $\gamma$ from the combination with the profile-likelihood (blue shaded region) and the plugin (black data points) methods, for all the $D$-decay modes. The fit result is $\gamma = (63 \pm 13)^{\circ}$.}
    \label{fig_combo_1d_gamma_all}
\end{figure}

\begin{alignat*}{4}
\gamma\phantom{_{B}^{DK^{*\pm}}} &= (63 \pm 13)^{\circ},\\
r_{B}^{DK^{*\pm}} &= 0.103 \pm 0.010,\\
\delta_{B}^{DK^{*\pm}} &= (47 ^{+14}_{-12})^{\circ}.
\end{alignat*}
The uncertainties are found to be the same with the ``plugin'' method.


 In summary, proton-proton collision data corresponding to an integrated luminosity of 9\invfb collected by the \lhcb experiment at centre-of-mass energies of $\sqrt{s}=7, \, 8$ and $13$\tev is used to perform 
a measurement of $\gamma$ using the decay \mbox{$\Bpm \to D \Kstarpm$}, where the $D$ meson decays to two-, three-, and four-body final states.
The measured value is \mbox{$\gamma = (63\pm 13)^\circ$}, where the uncertainty is statistically dominated. This measurement is a valuable addition to the knowledge of $\gamma$. The \CP-violating observables measured here are consistent with and supersede those presented in Ref.~\cite{LHCb-PAPER-2017-030}, and this measurement constitutes the first observation of the suppressed \mbox{$\Bpm \to [\pipm\Kmp]_D \Kstarpm$} and \mbox{$\Bpm \to [\pipm\Kmp\pipm\pimp]_D \Kstarpm$} decays. 

\section*{Acknowledgements}
%
%
\noindent We express our gratitude to our colleagues in the CERN
accelerator departments for the excellent performance of the LHC. We
thank the technical and administrative staff at the LHCb
institutes.
We acknowledge support from CERN and from the national agencies:
CAPES, CNPq, FAPERJ and FINEP (Brazil); 
MOST and NSFC (China); 
CNRS/IN2P3 (France); 
BMBF, DFG and MPG (Germany); 
INFN (Italy); 
NWO (Netherlands); 
MNiSW and NCN (Poland); 
MCID/IFA (Romania); 
MICIU and AEI (Spain);
SNSF and SER (Switzerland); 
NASU (Ukraine); 
STFC (United Kingdom); 
DOE NP and NSF (USA).
We acknowledge the computing resources that are provided by CERN, IN2P3
(France), KIT and DESY (Germany), INFN (Italy), SURF (Netherlands),
PIC (Spain), GridPP (United Kingdom), 
CSCS (Switzerland), IFIN-HH (Romania), CBPF (Brazil),
and Polish WLCG (Poland).
We are indebted to the communities behind the multiple open-source
software packages on which we depend.
Individual groups or members have received support from
ARC and ARDC (Australia);
Key Research Program of Frontier Sciences of CAS, CAS PIFI, CAS CCEPP, 
Fundamental Research Funds for the Central Universities, 
and Sci. \& Tech. Program of Guangzhou (China);
Minciencias (Colombia);
EPLANET, Marie Sk\l{}odowska-Curie Actions, ERC and NextGenerationEU (European Union);
A*MIDEX, ANR, IPhU and Labex P2IO, and R\'{e}gion Auvergne-Rh\^{o}ne-Alpes (France);
AvH Foundation (Germany);
ICSC (Italy); 
Severo Ochoa and Mar\'ia de Maeztu Units of Excellence, GVA, XuntaGal, GENCAT, InTalent-Inditex and Prog. ~Atracci\'on Talento CM (Spain);
SRC (Sweden);
the Leverhulme Trust, the Royal Society
 and UKRI (United Kingdom).



\newpage
\section*{Correlation matrices}

\begin{table}[h!]
\caption{Correlation matrix of statistical uncertainties on $\CP$ observables for two- and four-body modes. Values below 0.1\% are indicated with the symbol ``--''.}
\resizebox{\textwidth}{!}{%
\begin{tabular}{ccccccccccccc}
\hline
\hline \\[0.005ex]
 & $A_{SS}^{K\pi}$ & $A_{\CP}^{KK}$ & $A_{\CP}^{\pi \pi}$ & $A_{OS}^{\pi K}$ & $R_{\CP}^{KK}$ & $R_{\CP}^{\pi \pi}$ & $R_{OS}^{\pi K}$ & $A_{SS}^{K\pi\pi\pi}$ & $A_{\CP}^{\pi \pi \pi \pi}$ & $A_{OS}^{\pi K \pi \pi}$ & $R_{\CP}^{\pi \pi \pi \pi}$ & $R_{OS}^{\pi K \pi \pi}$ \\[1ex]
\hline
$A_{SS}^{K\pi}$    & $1$ & -- & -- & -- & -- & -- & -- & -- & -- & -- & -- & -- \\[1ex]
$A_{\CP}^{KK}$      & & $1$ & -- & -- & $-0.019$ & -- & -- & -- & -- & -- & -- & -- \\[1ex]
$A_{\CP}^{\pi \pi}$ & & & $1$ & -- & -- & $-0.085$ & -- & -- & -- & -- & -- & -- \\[1ex]
$A_{OS}^{\pi K}$   & & & & $1$ & -- & -- & $0.183$ & -- & -- & -- & -- & -- \\[1ex]
$R_{\CP}^{KK}$      & & & & & $1$ & $~~0.055$ & $0.028$ & -- & -- & -- & -- & -- \\[1ex]
$R_{\CP}^{\pi \pi}$ & & & & & & $1$ & $0.020$ & -- & -- & -- & -- & -- \\[1ex]
$R_{OS}^{\pi K}$   & & & & & & & $1$ & -- & -- & -- & -- & -- \\[1ex]
$A_{SS}^{K\pi\pi\pi}$      & & & & & & & & $1$ & -- & -- & -- & -- \\[1ex]
$A_{\CP}^{\pi \pi \pi \pi}$ & & & & & & & & &$1$ & -- & $-0.022$& -- \\[1ex]
$A_{OS}^{\pi K \pi \pi}$   & & & & & & & & & & $1$ & -- & $0.040$ \\[1ex]
$R_{\CP}^{\pi \pi \pi \pi}$ & & & & & & & & & & & $1$ & $0.029$ \\[1ex]
$R_{OS}^{\pi K \pi \pi}$   & & & & & & & & & & & & $1$ \\[1ex]
\hline
\hline
\end{tabular}
}
\label{Table:Mcorr_stat_adsglw}
\end{table}

\begin{table}[h!]
\caption{Correlation matrix of systematic uncertainties on $\CP$ observables for two- and four-body modes. Values below 0.1\% are indicated with the symbol ``--''.}
\resizebox{\textwidth}{!}{%
\begin{tabular}{ccccccccccccc}
\hline 
\hline \\[0.005ex]
 & $A_{SS}^{K\pi}$ & $A_{\CP}^{KK}$ & $A_{\CP}^{\pi \pi}$ & $A_{OS}^{\pi K}$ & $R_{\CP}^{KK}$ & $R_{\CP}^{\pi \pi}$ & $R_{OS}^{\pi K}$ & $A_{SS}^{K\pi\pi\pi}$ & $A_{\CP}^{\pi \pi \pi \pi}$ & $A_{OS}^{\pi K \pi \pi}$ & $R_{\CP}^{\pi \pi \pi \pi}$ & $R_{OS}^{\pi K \pi \pi}$ \\[1ex]
\hline
$A_{SS}^{K\pi}$    & $1$ & $0.314$ & -- & $-0.014$ & -- & -- & $-0.016$ & $~~0.918$ & $0.010$ & $-0.169$ & -- & -- \\[1ex]
$A_{\CP}^{KK}$      & & $1$ & -- & $-0.017$ & $-0.039$ & -- & $-0.011$ & $~~0.319$ & $0.032$ & $~~0.155$ & $-0.019$ & $0.048$ \\[1ex]
$A_{\CP}^{\pi \pi}$ & & & $1$ & -- & -- & $-0.292$ & -- & -- & -- & -- & -- & -- \\[1ex]
$A_{OS}^{\pi K}$   & & & & $1$ & -- & -- & $~~0.220$ & $-0.017$ & -- & $~~0.086$ & -- & $0.023$ \\[1ex]
$R_{\CP}^{KK}$      & & & & & $1$ & $~~0.051$ & -- & -- & -- & -- & -- & -- \\[1ex]
$R_{\CP}^{\pi \pi}$ & & & & & & $1$ &-- & -- & -- & -- & -- & -- \\[1ex]
$R_{OS}^{\pi K}$   & & & & & & & $1$ & -- & -- & $-0.013$ & -- & $0.063$ \\[1ex]
$A_{SS}^{K\pi\pi\pi}$      & & & & & & & & $1$ & $0.013$ & $-0.173$ & -- & $0.013$ \\[1ex]
$A_{\CP}^{\pi \pi \pi \pi}$ & & & & & & & & &$1$ & -- & $-0.605$ & $0.025$ \\[1ex]
$A_{OS}^{\pi K \pi \pi}$   & & & & & & & & & & $1$ & -- & $0.345$ \\[1ex]
$R_{\CP}^{\pi \pi \pi \pi}$ & & & & & & & & & & & $1$ & -- \\[1ex]
$R_{OS}^{\pi K \pi \pi}$   & & & & & & & & & & & & $1$ \\[1ex]
\hline
\hline
\end{tabular}
}
\label{Table:Mcorr_syst_adsglw}
\end{table}

\begin{table}[h!]
\caption{Correlation matrix of statistical uncertainties on $\CP$ observables for three-body modes. Values below 0.1\% are indicated with the symbol ``--''.}
\centering
\begin{tabular}{ccccc}
\hline
\hline
 & $x_+$ & $y_+$ & $x_-$ & $y_-$ \\
\hline
$x_+$  & $1$ & $-0.144$ & -- & -- \\
$y_+$  & & $1$ & -- & -- \\
$x_-$  & & & $1$ & $0.448$       \\
$y_-$  & & & & $1$          \\
\hline
\hline
\end{tabular}
\label{Table:Mcorr_stat_bpggsz}
\end{table}

\begin{table}[h!]
\caption{Correlation matrix of systematic uncertainties on $\CP$ observables for three-body modes.}
\centering
\begin{tabular}{ccccc}
\hline
\hline
 & $x_+$ & $y_+$ & $x_-$ & $y_-$ \\
\hline
$x_+$  & $1$ & $-0.055$ & $~~0.030$ & $~~0.022$ \\
$y_+$  & & $1$ & $-0.158$ & $-0.478$ \\
$x_-$  & & & $1$ & $~~0.319$       \\
$y_-$  & & & & $1$          \\
\hline
\hline
\end{tabular}
\label{Table:Mcorr_syst_bpggsz}
\end{table}

\begin{table}
\caption{Correlations in the $\CP$ observables for the strong-phase related systematic uncertainties in three-body modes of different $B$-meson decay channels. The blocks of rows correspond to $\btodh$~\cite{LHCb-PAPER-2020-019}, $\btodkstarz$~\cite{LHCb-PAPER-2023-009}, fully reconstructed $\btodstarh$~\cite{LHCb-PAPER-2023-012}, partially reconstructed $\btodstarh$~\cite{LHCb-PAPER-2023-029} and $\btodkstar$ (this analysis) channels, respectively.}
\centering
    \begin{tabular}{ l c c c c }

\hline
\hline

 &  $x_{-}^{DK^{*+}}$ & $y_{-}^{DK^{*+}}$ & $x_{+}^{DK^{*+}}$ & $y_{+}^{DK^{*+}}$ \\
\hline

$x_{-}^{DK}$ & $-0.269$ & $-0.073$ & $~~0.196$ & $-0.039$ \\
$y_{-}^{DK}$ & $-0.528$ & $-0.647$ & $-0.241$ & $~~0.457$ \\
$x_{+}^{DK}$ & $-0.140$ & $-0.003$ & $-0.157$ & $~~0.010$ \\
$y_{+}^{DK}$ & $-0.422$ & $-0.248$ & $~~0.267$ & $-0.082$ \\
$\Re(\xi^{D\pi})$ & $~~0.013$ & $~~0.275$ & $-0.078$ & $-0.278$ \\
$\Im(\xi^{D\pi})$ & $~~0.305$ & $~~0.533$ & $~~0.322$ & $-0.534$ \\
\hline 
$x_{-}^{D\Kstarz}$ & $-0.034$ & $-0.068$ & $-0.024$ & $~~0.076$ \\
$y_{-}^{D\Kstarz}$ & $-0.028$ & $-0.038$ & $-0.015$ & $~~0.014$ \\
$x_{+}^{D\Kstarz}$ & $~~0.052$ & $~~0.056$ & $~~0.094$ & $-0.033$ \\
$y_{+}^{D\Kstarz}$ & $~~0.033$ & $~~0.010$ & $~~0.007$ & $-0.015$ \\
\hline
$x_{-}^{\Dstar K}$ full reco & $-0.047$ & $-0.051$ & $~~0.010$ & $~~0.023$ \\
$y_{-}^{\Dstar K}$ full reco & $~~0.022$ & $~~0.022$ & $-0.020$ & $-0.013$ \\
$x_{+}^{\Dstar K}$ full reco & $-0.033$ & $-0.046$ & $~~0.009$ & $~~0.042$ \\
$y_{+}^{\Dstar K}$ full reco & $-0.025$ & $-0.048$ & $~~0.029$ & $~~0.024$ \\
$\Re(\xi^{\Dstar \pi})$ full reco & $~~0.059$ & $~~0.056$ & $~~0.002$ & $-0.022$ \\
$\Im(\xi^{\Dstar \pi})$ full reco & $~~0.046$ & $~~0.028$ & $-0.013$ & $~~0.011$ \\
\hline
$x_{-}^{\Dstar K}$ part reco & $-0.039$ & $-0.012$ & $~~0.032$ & $~~0.041$ \\
$y_{-}^{\Dstar K}$ part reco & $-0.026$ & $-0.005$ & $~~0.020$ & $~~0.013$ \\
$x_{+}^{\Dstar K}$ part reco & $~~0.008$ & $~~0.035$ & $-0.040$ & $~~0.002$ \\
$y_{+}^{\Dstar K}$ part reco & $-0.024$ & $-0.053$ & $~~0.006$ & $~~0.005$ \\
$\Re(\xi^{\Dstar \pi})$ part reco & $-0.042$ & $-0.018$ & $~~0.013$ & $~~0.020$ \\
$\Im(\xi^{\Dstar \pi})$ part reco & $-0.044$ & $-0.008$ & $~~0.000$ & $~~0.026$ \\
\hline
 $x_{-}^{DK^{*+}}$ & $~~1.000$ & $-0.520$ & $~~0.080$ & $~~0.170$\\
 $y_{-}^{DK^{*+}}$ &    & $~~1.000$ & $-0.410$ & $-0.560$\\
 $x_{+}^{DK^{*+}}$ &    &    & $~~1.000$ & $~~0.760$\\
 $y_{+}^{DK^{*+}}$ &    &    &    & $~~1.000$\\
\hline
\hline
\end{tabular}
\label{Table:cisicorr}
\end{table}

\clearpage

\appendix


\addcontentsline{toc}{section}{References}
\bibliographystyle{LHCb}
\bibliography{main,standard,LHCb-PAPER,LHCb-CONF,LHCb-DP,LHCb-TDR}

\newpage
\centerline
{\large\bf LHCb collaboration}
\begin
{flushleft}
\small
R.~Aaij$^{37}$\lhcborcid{0000-0003-0533-1952},
A.S.W.~Abdelmotteleb$^{56}$\lhcborcid{0000-0001-7905-0542},
C.~Abellan~Beteta$^{50}$,
F.~Abudin{\'e}n$^{56}$\lhcborcid{0000-0002-6737-3528},
T.~Ackernley$^{60}$\lhcborcid{0000-0002-5951-3498},
A. A. ~Adefisoye$^{68}$\lhcborcid{0000-0003-2448-1550},
B.~Adeva$^{46}$\lhcborcid{0000-0001-9756-3712},
M.~Adinolfi$^{54}$\lhcborcid{0000-0002-1326-1264},
P.~Adlarson$^{81}$\lhcborcid{0000-0001-6280-3851},
C.~Agapopoulou$^{14}$\lhcborcid{0000-0002-2368-0147}
% ...

{\footnotesize \it

% ...
$^{14}$Universit{\'e} Paris-Saclay, CNRS/IN2P3, IJCLab, Orsay, France\\
% ...
$^{37}$Nikhef National Institute for Subatomic Physics, Amsterdam, Netherlands\\
% ...
$^{46}$Instituto Galego de F{\'\i}sica de Altas Enerx{\'\i}as (IGFAE), Universidade de Santiago de Compostela, Santiago de Compostela, Spain\\
% ...
$^{50}$Physik-Institut, Universit{\"a}t Z{\"u}rich, Z{\"u}rich, Switzerland\\
% ...
$^{54}$H.H. Wills Physics Laboratory, University of Bristol, Bristol, United Kingdom\\
% ...
$^{56}$Department of Physics, University of Warwick, Coventry, United Kingdom\\
% ...
$^{59}$School of Physics and Astronomy, University of Glasgow, Glasgow, United Kingdom\\
$^{60}$Oliver Lodge Laboratory, University of Liverpool, Liverpool, United Kingdom\\
% ...
$^{68}$Syracuse University, Syracuse, NY, United States\\
% ...
$^{81}$Department of Physics and Astronomy, Uppsala University, Uppsala, Sweden, associated to $^{59}$\\
% ...
}
\end{flushleft}




\end{document}